\documentclass[trackchanges,twocolumn]{aastex701}

\usepackage{amsmath}
\usepackage{float}
\usepackage{xcolor}
\usepackage{booktabs,tabularx,array}

\begin{document}

\title{Dipole Anisotropy in Newly Sampled AGN Population: Still in Tension with CMB Dipole} 

\author[orcid=0009-0008-0618-6623,sname='Lee']{Dahee Lee}
\affiliation{Department of Earth Sciences, Chosun University, Jeonnam-Gwangju 61452, Republic of Korea}
\email[show]{daheelee@chosun.ac.kr}  

\author[orcid=0000-0003-4181-138X,sname='Oh']{Minji Oh} 
\affiliation{Department of Earth Sciences, Chosun University, Jeonnam-Gwangju 61452, Republic of Korea}
\email{minji.wow@gmail.com}

\author[orcid=0009-0001-5284-6759,sname='Asok']{Gayathri Asok}
\affiliation{{Division of Particle and Astrophysical Science, Nagoya University, Furo-cho, Chikusa-ku, Nagoya 464-8602, Japan}}
\email{t.gayathriasok@gmail.com}

\author[orcid=0000-0001-8416-7673,sname='Takeuchi']{Tsutomu T.\ Takeuchi}
\affiliation{{Division of Particle and Astrophysical Science, Nagoya University, Furo-cho, Chikusa-ku, Nagoya 464-8602, Japan}}
\affiliation{{The Research Center for Statistical Machine Learning, the Institute of Statistical Mathematics, 10--3 Midori-cho, Tachikawa, Tokyo 190--8562, Japan}}
\email{tsutomu.takeuchi.ttt@gmail.com}

\author[orcid=0000-0003-3974-1239,sname='Ahn']{Kyungjin Ahn}
\affiliation{Department of Earth Sciences, Chosun University, Jeonnam-Gwangju 61452, Republic of Korea}
\email[show]{kjahn@chosun.ac.kr}

\correspondingauthor{Kyungjin Ahn}

\begin{abstract}

We revisit the all-sky survey data, the Wide-field Infrared Survey Explorer (WISE) catalog, with an improved galaxy-selection criterion to investigate the inconsistency between the dipole anisotropy in the galaxy number count and the kinematic CMB dipole anisotropy. We apply a set of active galactic nuclei (AGN) selection criteria composed of a flux cut and multiple color cuts, and formulate an appropriate adaptation of the original Ellis \& Baldwin (1984) formalism for our photometric survey samples whose spectral energy distributions deviate from a pure power-law form. We find that the newly sampled AGNs from the WISE catalog still bear an anomalous number-count dipole anisotropy, whose amplitude is $\sim 2.5$ (2.1) times as large as that of the purely kinematic origin at a $5.4\sigma$  (3.9$\sigma$) confidence level if the apparent ecliptic bias is uncorrected (corrected). We also find that the direction of our WISE-AGN dipole is to occur only {at $\sim1.1\%$ probability} if the dipole is of purely kinematic origin, whether or not the ecliptic bias is corrected. We discuss its possible cosmological implications {and systematic uncertainties}: nonlinearity of the nearby universe, mismatch of matter-rest frame and the CMB-rest frame{, and the systematic uncertainties that could impact the dipole estimate}.

\end{abstract}

\keywords{\uat{Observational cosmology}{1146} --- \uat{Cosmic
    microwave background radiation}{322} --- \uat{Large-scale
    structure of the universe}{902}}

\section{Introduction}
Modern cosmology is based on the assumption of a homogeneous and isotropic universe, known as the Cosmological Principle (CP). In the era of precision cosmology, one of the strongest pieces of evidence for isotropy is the cosmic microwave background radiation (CMB). The CMB appears isotropic and homogeneous on very large scales, with the monopole temperature $\left<T_{\rm CMB}\right>=2.725$~K \citep{Fixsen2009}. However, this isotropy is not absolute and tiny temperature fluctuations, or anisotropies, are imprinted across the sky. 

The most prominent of these is the largest-scale deviation, or the dipole anisotropy. After the first detection of the CMB dipole many decades ago \citep{Conklin1969}, subsequent observations have improved the accuracy substantially with an almost converging amplitude at $D=3.4$~mK (with $\Delta T(\theta)=D\cos \theta$ where $\theta$ is the azimuthal angle against the direction of the dipole) and its direction  $(l,b)=(264^\circ, 48^\circ)$ in Galactic coordinates \citep{Fixsen1996,WMAP2009,Planck2020}. The CMB dipole is widely attributed not to a primordial origin but to the motion of the observer. This indicates that the observer moves with respect to the CMB-rest frame with the solar velocity $v_\odot =370 $~km/s, or $\beta_\odot \equiv v_\odot /c=1.23 \times 10^{-3}$\citep{Fixsen1996} such that $D=\left<T_{\rm CMB}\right> \beta_\odot$. The present $1\sigma$ upper bound on the intrinsic CMB dipole is roughly $3.7$~mK, which is similar in magnitude to the measured CMB dipole \citep{Ferreira2021}. Even so, the measured dipole is far larger than the higher-order multipoles, making a kinematic origin the most straightforward interpretation.

If the matter-rest frame and the CMB-rest frame coincide with each other as expected \citep{Ehlers1968,Stoeger1995}, a flux-limited survey of astrophysical sources should also exhibit a dipole anisotropy of the same origin ($v_\odot$) due to the Doppler effect and the angular aberration. \citet[EB hereafter]{EB84} first formulated the idea of the dipole anisotropy in an astrophysical source distribution, applicable to continuum-emitting galaxies. Applying a universal flux cut within a fixed solid angle when sampling galaxies in the observing frame induces a non-universal sampling of galaxies in the matter-rest frame and thus the dipole anisotropy. However, previous studies have tended to indicate that the galactic dipole anisotropy is too large to be attributed to the same solar velocity \citep{Singal2011,Secrest2021,Wagenveld2023}, with the dipole direction sometimes marginally consistent with the CMB one depending on specific survey data \citep{Secrest2021,Darl22,Wagenveld2023}. This is known as the ``cosmic dipole tension'', which remains one of the unsolved problems in cosmology\footnote{See however \citet{Darl22} claiming that the two dipoles are consistent with each other, based on the combined data of VLASS and the Rapid Australian Square Kilometer Array Pathfinder Continuum Survey (RACS). \citet{Darl22} ascribes the dipole tension to an artifact arising from systematic uncertainties. A similar claim also comes from the dipole in a galaxy-redshift survey, or extended Baryon Oscillation Spectroscopic Survey (eBOSS; \citealt{daSilvFerr2024}), yet the consensus in redshift-survey dipoles is still against such a consistency.}. Several studies using various radio-galaxy survey data, such as the NRAO VLA Sky Survey (NVSS), the Very Large Array Sky Survey (VLASS), the WEsterbork Northern Sky Survey (WENSS), the TIFR GMRT Sky Survey (TGSS), and the Rapid Australian Square Kilometer Array Pathfinder Continuum Survey (RACS) have reported this dipole anomaly with estimated $v_\odot$ which is $\sim2-5$ times as large as that from the CMB dipole with high significance, on the order of $\gtrsim[3-5]\sigma$ \citep{Singal2011,Rubart2013,Bengaly2018,Singal2024}.
The same type of anomaly is observed also in another spectral range: in infrared, the Wide-field Infrared Survey Explorer (WISE) data shows a dipole anisotropy that is too large to be accounted for by $v_\odot\sim370$~km/s \citep{Rameez2018,Secrest2021,Dam2023}.

Any resolution to the dipole tension falls into one of the following three categories: (1) the matter-rest frame is moving against the CMB-rest frame, (2) we are in a rare location in the Universe with unusually large intrinsic inhomogeneity around, and (3) the systematic uncertainties in galaxy surveys are substantial enough to produce a spurious dipole anisotropy. The first resolution is against the current standard model of the Universe. The second resolution is highly unlikely from a statistical standpoint, if not impossible \citep{Tiwari2016,Bashir2026}. Although the third resolution may be the most ordinary, the prevailing consensus so far still indicates a significant degree of tension.

Independent probes, careful treatment of any systematic uncertainties in survey data and survey-specific formulation of the dipole anisotropy would help either improving the statistical significance of the dipole tension or easing the tension. \citet{Maartens2018} and \citet{Ahn2025} suggested observing dipole anisotropy in line intensity mapping (LIM) surveys\footnote{\citet{Maartens2018} is the first one to investigate a dipole in the diffuse 21~cm LIM (see also e.g. \citealt{Hotinli2024}), and \citet{Ahn2025} considered both the diffuse LIM dipole and the number count dipole of line-emitting galaxies.} instead of continuum-source surveys. The kinematic dipole can show up in both the 1-point statistics of spectroscopic redshift surveys via source number counts (\citealt{Guandalin2023,Tiwari2024,daSilvFerr2024,vonHause2025}) and the 2-point clustering statistics through the galaxy correlation function (\citealt{Bonvin2014, Raccanelli2018, Lacasa2024, TakeuchiB2026}). Survey-specific interpretation of the dipole amplitude requires generalization of the original EB formula, such as the large-bandwidth impact in photometry and the breakdown of power-law assumptions \citep{TakeuchiA2026,Bonnefous2026}. 

We revisit the All WISE Data Release (AllWISE: \citealt{AllWISE,AllWISE_catalog}), the full data product of the WISE survey, and calculate the number-count dipole anisotropy to assess the dipole tension with yet another new set of AGNs. This will provide another (almost) independent test of the dipole tension. In order to select AGNs from the data, we use the three-color-cut criterion by \citet{Blecha2018}. Revisiting the AllWISE dataset with this well-tested selection criterion provides a strategic advantage for constraining the cosmic dipole anisotropy. Quantifying this small dipole anisotropy requires a large number of samples to minimize the shot noise \citep{Maartens2018}, and at the same time contamination by e.g. stellar objects should be minimized. Standard single-color thresholds, such as the $W1-W2 > 0.8$ cut by \citet{Stern2012}, are known to suffer from stellar contamination. Conversely, over-correcting with highly restrictive cuts discards lower-luminosity and heavily obscured AGNs, weakening the statistical power of the survey. By adopting the \citet{Blecha2018} selection framework, we optimize this trade-off: it efficiently isolates and mitigates stellar contaminants while simultaneously recovering the obscured and host-diluted AGNs missed by rigid single-color cuts. This approach yields a large number of samples still comparable to those from \citet{Stern2012} yet with presumably better selection of AGN systems.

Through the process of our calculating the dipole anisotropy in the new catalog of AGNs, we realized the necessity to reformulate the EB formula suitable for color-cut applied dipole measure. While very limited in the number of color indices, the photometric color indices in our AGN catalog samples revealed that the mean spectral energy distribution (SED) deviates from a simple power-law shape. This then generates another type of dipole anisotropy, or ``the color-cut dipole anisotropy'', which we formulate and apply in this paper (for a similar generalization to such a color-cut dipole anisotropy, see also \citealt{vonHause2025}). We will show that this gives a substantial modification to the original EB formula, yet limited to a change of the dipole amplitude at $\lesssim 10\%$ level in our case.


This paper is organized as follows. We mathematically formulate the kinematic dipole anisotropy in the number count of continuum sources selected with both a flux-cut and multi-color-cuts, and also show our scheme to estimate the dipole anisotropy from a number-count map in Section~\ref{sec:dipole}. We brief the WISE data product in Section~\ref{sec:data} and describe the details of our data-reduction and sample-selection processes in Section~\ref{sec:sample}. Finally, we report and discuss our results in Sections~\ref{sec:results} and \ref{sec:discussion}, respectively.
\section{Dipole anisotropy} \label{sec:dipole}
\subsection{Ellis-Baldwin formalism with Jacobians} \label{subsec:EB}
We re-derive the formula (``EB formula'') for the dipole anisotropy in a flux-limited number count of power-law continuum sources derived in EB but in a slightly different manner. As long as the field-defining variables are explicitly treated, the dipole anisotropy is derived from Jacobians and Taylor expansions. While trivial, in Section~\ref{subsec:colorcut} this procedure will serve as a proxy for our own modification of the EB formula that is required for a \emph{photometric survey} with large frequency bandwidths and the flux \emph{spectrum under a broken power-law}.

The relativistic Doppler factor $\delta$ is given by
\begin{equation}
    \delta = \frac{1}{\gamma(1-\beta \cos \theta')},
\end{equation}
where $\gamma = (1-v^2/c^2)^{-1/2}$ is the Lorentz factor, $\beta=v/c$ is the velocity of the observer as a fraction of the speed of light, and $\theta'$ denotes the angle between the velocity vector and the line of sight in the observer-rest frame (ORF). 
Due to the Doppler effect, the following transformation rules between the two frames occur \citep[e.g.][]{Ahn2025}:
\begin{equation}
    \nu' = \nu \delta, \ d\Omega'=d\Omega \delta^{-2}, \ \cos\theta' = \frac{\cos\theta + \beta}{1+\beta \cos\theta},
\label{eq:transform}
\end{equation}
where $\Omega$ is the solid angle, and the primed variables denote the quantities in ORF and un-primed variables those in the matter-rest frame (MRF). 

For a population that exhibits a power-law spectral energy distribution, the flux density $S$ ($\mathrm{erg\,s^{-1}\,cm^{-2}\,Hz^{-1}}$) at frequency $\nu$ is expressed as 
\begin{equation}
     S \propto \nu^{-\alpha}.
     \label{eq:flux}
\end{equation}
The Doppler effect relates the observed $S'$ to the rest-frame $S$ by
\begin{equation}
    S'(\nu') = S(\nu)\delta,
\end{equation}
and this leads to $S'(\nu') = S(\nu') \delta^{1+\alpha}$ from Eqs.~(\ref{eq:transform}) and (\ref{eq:flux}).

The differential number density $n(S)$ is defined as
\begin{equation}
    n(S) = \frac{dN}{dS \ d\Omega},
\end{equation}
and the cumulative number density above a limiting flux density $\bar{S}$ per unit solid angle is given by
\begin{equation}
    \mathcal{N}(>\bar{S})=\frac{dN}{d\Omega}(>\bar{S})=\int_{\bar{S}} dS \ n(S).
\end{equation}
Since the Lagrangian number is conserved ($dN'=dN$), we have
\begin{equation}\label{eq:dN}
    \mathcal{N'}(>\bar{S}')=\mathcal{N}(>\bar{S}) \frac{d\Omega}{d\Omega'}=\mathcal{N}(>\bar{S})\delta^2.
\end{equation}
Using a first-order Taylor expansion of $\mathcal{N}(>\bar{S})$ about an ORF variable $\bar{S}'$ and Eq.~\ref{eq:dN}, the amplitude of the dipole moment of $\mathcal{N'}$ (we take the convention that $\mathcal{N'}(\theta')=\mathcal{N'}_{\rm dip}\cos\theta'$) is obtained as
\begin{equation}
    \mathcal{N'}_\text{dip}(>\bar{S}') = \mathcal{N}(>\bar{S}') \left( 2-(1+\alpha) \left. \frac{\partial \ln\mathcal{N}}{\partial \ln \bar{S}}\right\vert_{\bar{S}'}\right)\beta.
\end{equation}
Here, $-\left. \partial \ln\mathcal{N}/\partial \ln \bar{S}\right\vert_{\bar{S}'}$ corresponds to $x$ in the EB formula. Therefore, the dimensionless dipole amplitude becomes
\begin{equation}
    \mathcal{D}\equiv\frac{\mathcal{N'}_\text{dip}(>\bar{S}')}{\mathcal{N}(>\bar{S}')}= \left[ 2+x(1+\alpha) \right]\beta,
\end{equation}
as given in EB.


\subsection{Dipole from color cuts: modification to Ellis-Baldwin formula} \label{subsec:colorcut}
A color-cut-induced dipole does not arise in a single-power-law SED since the Doppler-boosted flux does not change the observed spectral index from the MRF value $\alpha$. For a more realistic SED such as a broken-power-law SED, however, the band-averaged flux in a band that contains the ``break frequency'', where the spectral power index changes, generates a dipole anisotropy in the observed color index. This happens because the Doppler shift of the break frequency alters the relative contributions of the two spectral slopes within the band, and the overall effect is a nontrivial Doppler boost of the band-averaged flux density (Fig.~\ref{fig:illust}; see details in \hyperref[app:colordipole]{Appendix}).

When a color index is measured using band-averaged fluxes in this band and an adjacent one, the resulting color index will depend on the line-of-sight direction. Consequently, a ``color-index dipole'' occurs. The color index in ORF can thus be written as a function of $\delta$ and expanded, to first order in $\ln\delta$, as 
\begin{align}
    C'(\delta) &= C'(1) + \left.\frac{dC'}{d\ln\delta}\right\vert_{\delta=1} \ln\delta.
\end{align}
For galaxies selected with color-index threshold, $C'>\bar{C}'$, the dipole moment of $\mathcal{N}$ induced by the combined flux and color cuts is given by
\begin{align}\label{eq:colordipole}
    &\mathcal{N'}_\text{dip}(>\bar{S}',>\bar{C}') \nonumber \\
    & \quad = \mathcal{N}(>\bar{S}',>\bar{C}') \left( 2+x(1+\alpha) +x_C\left.\frac{dC'}{d\ln\delta}\right\vert_{\delta=1}\right)\beta.
\end{align}
where 
\begin{align*}
    x &\equiv \left.-\frac{\partial \ln\mathcal{N}}{\partial \ln \bar{S}}\right|_{\bar{S}',\,\bar{C}'}, &
    x_C &\equiv \left.-\frac{\partial \ln\mathcal{N}}{\partial \bar{C}}\right|_{\bar{S}',\,\bar{C}'}.
\end{align*}

Since multiple color cuts are applied in this work, Eq.~\ref{eq:colordipole} can be generalized to account for multiple effects, and the resulting expression for the dipole takes the following final form (see an explicit expression and the resulting dipole amplitude in Appendix~\ref{app:colordipole}):
\begin{align}\label{eq:multicolordipole}
    &\mathcal{N'}_\text{dip}(>\bar{S}',>\bar{C}_{1}',>\bar{C}_{2}',>\bar{C}_{3}',...) \nonumber \\
    & = \mathcal{N}(>\bar{S}',>\bar{C}_{1}',>\bar{C}_{2}',>\bar{C}_{3}',...) \nonumber \\
    &\quad\left( 2+x(1+\alpha) +\sum_i x_{C_i}\left.\frac{dC_{i}'}{d\ln\delta}\right\vert_{\delta=1}\right)\beta,
\end{align}
where $x_{C_i}$ is defined as a logarithmic change rate of the cumulative number density $\mathcal{N}$ with respect to $C_i$ but with all other color-index cuts and the flux cut fixed, or
\begin{equation*}
x_{C_i} \equiv -\left.\frac{\partial \ln\mathcal{N}}{\partial \bar{C_i}}\right|_{\bar{S}',\,\bar{C_j},\,\bar{C_k},...}
\end{equation*}
where $i\notin$\{$j$, $k$,...\}. We use three color cuts made of $C_1 \equiv W1 -W2$, $C_2 \equiv W2 - W3$ and a derived color index $C_3 \equiv C_1 -2C_2$, following \citet{Blecha2018}: $C_1 > \bar{C'_1}=0.5$, $C_2 >\bar{C'_2}= 2.2$ and $C_3 >\bar{C'_3}= -8.9$.

\subsection{Estimation of the dipole anisotropy} \label{subsec:estdipole}
To determine the dipole anisotropy, we employ the least-squares fitting method. In {\fontfamily{qcr}\selectfont healpy}, a Python package for HEALPix \citep{healpy}, the {\fontfamily{qcr}\selectfont fit\_dipole} routine provides the best-fit values for the monopole and dipole of the given map. This estimator minimizes the following function:
\begin{equation}\label{eq:fit}
    \sum_i \left[ N_i-\bar{N}(1 + \vec{\mathcal{D}}\cdot \hat{d}_i)\right]^2,
\end{equation}
where $N_i$ is the number of samples in the $i$-th pixel, $\hat{d}_i$ is the directional unit vector of the pixel $i$, and $\bar{N}$ and $\vec{\mathcal{D}}$ denote the amplitudes of the monopole and dimensionless dipole, respectively. 
Not all the pixels are used, because we apply a mask (Section~\ref{sec:sample}). The unmasked pixels are equally weighted in Eq.~\ref{eq:fit}.

\section{Data: AllWISE} \label{sec:data}
\begin{figure}[ht!]
    \centering
    \includegraphics[width=0.99\linewidth]{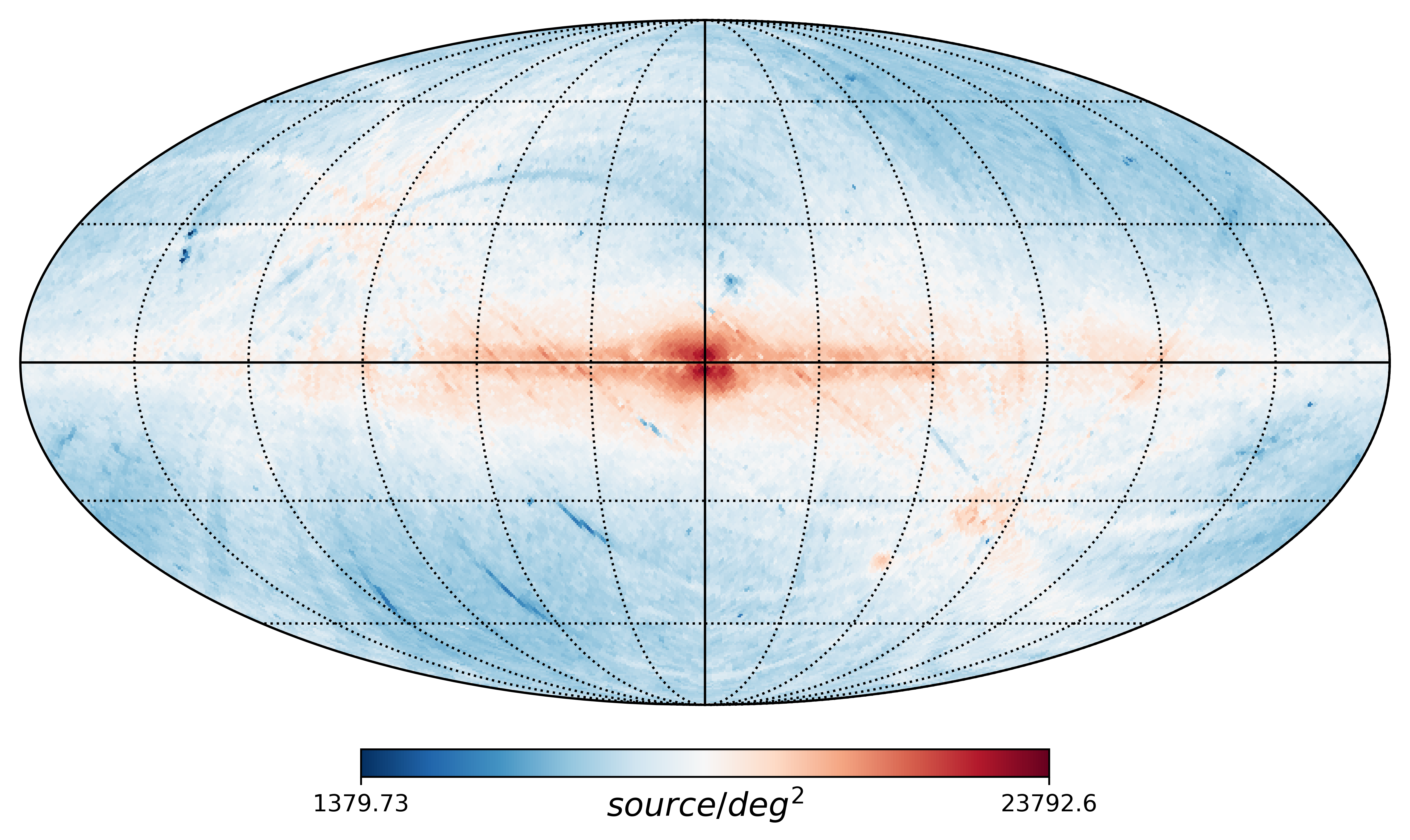}
    \caption{Number density map of the AllWISE catalog in Galactic coordinates. The color map represents the number of sources per square degree (NSIDE=64, or total 49152 pixels).}
    \label{fig:AW_numden}
\end{figure}
The Wide-field Infrared Survey Explorer (WISE) mission is an all-sky mid-infrared survey launched in 2009 \citep{WISE}. The survey was conducted in four bands, centered at 3.4, 4.6, 12, and 22 $\mu\text{m}$, corresponding to bands $W1$, $W2$, $W3$, and $W4$, respectively. The first full-sky survey data was released in 2012 through the WISE All-Sky Data Release, followed subsequently by the WISE 3-Band Cryo and NEOWISE Post-Cryo Data Releases. The AllWISE data release became available in 2013 by combining these previous data releases, containing 747,634,026 objects. It provides greater depth of coverage, sensitivity, and photometric accuracy than the WISE all-sky catalog due to the improved source extraction strategy \citep{AllWISE, AllWISE_catalog}. Fig.~\ref{fig:AW_numden} shows the number of AllWISE objects per HEALPix pixel.

The flux density in the AllWISE catalog is provided in units of Digital Number (DN). Hence, we convert the raw DN to Jansky (Jy) using the conversion coefficients of $1.9350\times 10^{-6}$, $2.7048\times 10^{-6}$, $2.9045\times 10^{-6}$, and $5.2269\times 10^{-5}$ for the $W1$, $W2$, $W3$, and $W4$ bands, respectively. To convert the flux density to magnitude, we use zero-magnitude flux densities of 309.540, 171.787, 31.674, and 8.363 Jy for the $W1$, $W2$, $W3$, and $W4$ bands, respectively \citep{AllWISE}.

\section{AGN selection}\label{sec:sample}
Our new catalog consists of AGNs selected from the AllWISE catalog with a flux cut and color cuts. Identifying AGNs in the mid-infrared is particularly useful because the dusty torus surrounding the central engine absorbs the high-energy radiation and re-emits it at infrared wavelengths. As a result, mid-infrared colors can provide an effective way to identify AGNs, including sources that are not detected at other wavelengths due to obscuration \citep{Goel2026}.

Before applying the color selection, we correct the WISE photometry for Galactic extinction to mitigate the effects of dust reddening. Following \citet{Wang2019}, we adopt extinction coefficients relative to the $V$ band, $A_{\lambda}/A_V$, of 0.039, 0.026, and 0.040 for $W1$, $W2$, and $W3$, respectively. For the color excess, $E(B-V)$, we use the Planck dust map provided through the {\fontfamily{qcr}\selectfont dustmaps} Python package \citep{Green2018, Planck2014}.

\begin{figure}[h]
    \centering
    \includegraphics[width=0.99\linewidth]{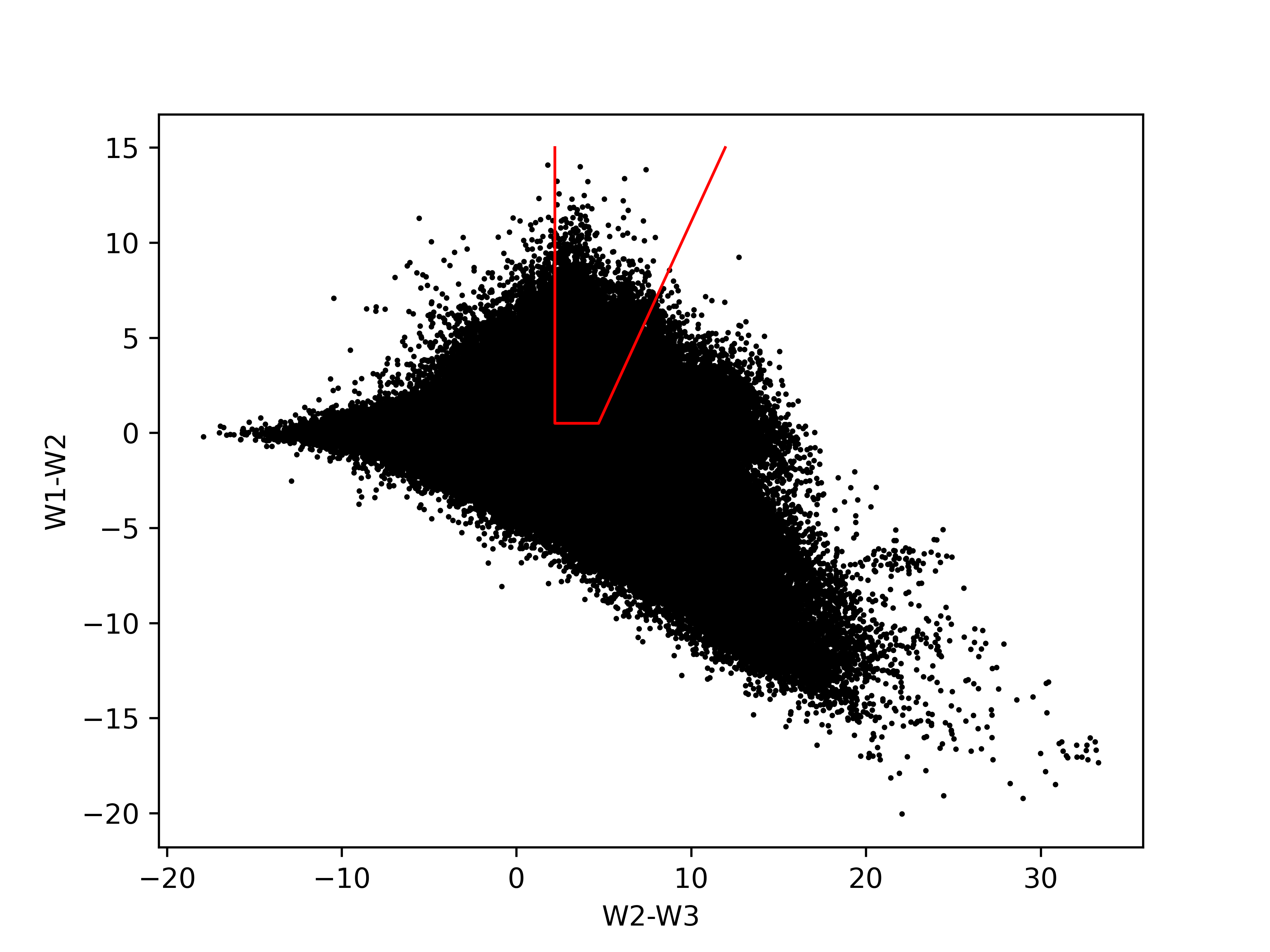}
    \caption{Color-color diagram of the AllWISE catalog. The red solid line denotes the AGN selection criteria proposed by \citet{Blecha2018}.}
    \label{fig:colorcut}
\end{figure}
We then apply the AGN 3-band color cut proposed by \citet{Blecha2018}:
\begin{align}
    [W1-W2] &> 0.5, \nonumber \\
    [W2-W3] &> 2.2, \nonumber \\
    [W1-W2] &> 2.0 \times [W2-W3] - 8.9.
\end{align}
We obtain 51,312,540 AGN candidates within the selection wedges in the color-color diagram of the AllWISE objects, as shown in Fig.~\ref{fig:colorcut}.


After applying the AGN selection criteria, we find spatial inhomogeneity in the source density map associated with the $W1$ magnitude limit. Due to the WISE scanning pattern, stripe-like artifacts appear repeatedly along the ecliptic latitude. Therefore, we also impose a magnitude cut on $W1$, $12 < W1 < 15.2$ in the Vega magnitude system or equivalently $0.26 < S_{W1} < 4.91$\,mJy in flux density, as suggested by \citet{Kovacs2015} and \citet{Rameez2018}. This leaves 1,270,202 AGN candidates.

\begin{figure}[ht!]
    \centering
    \includegraphics[width=0.99\linewidth]{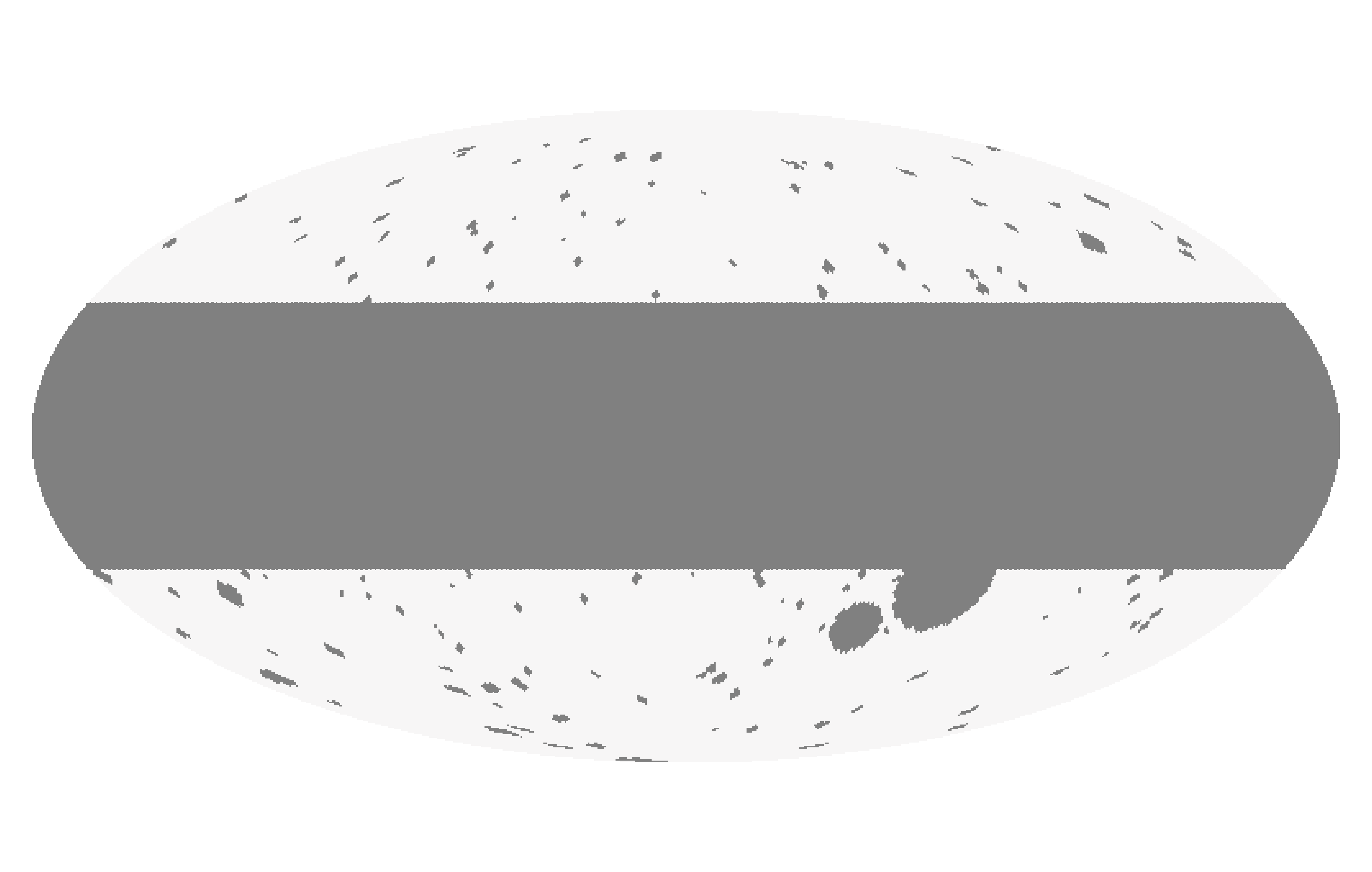}
    \caption{Sky mask applied in this work.}
    \label{fig:mask}
\end{figure}
We adopt the masking scheme used by \citet{Secrest2021}. Regions within a Galactic latitude of $|b|<30^\circ$ are masked to eliminate the strong foreground from the Galactic plane.
For nearby objects such as the Magellanic Clouds and M31, circular masks are applied. Additional circular masks based on the Two Micron All Sky Survey K-band-dependent radius relation $\log_{10}(r\ \mathrm{deg}^{-1}) = -0.134 K -0.471$ are also constructed to alleviate the effects of bright stars, whose image artifacts suppress the detection of nearby sources. Lastly, we mask the 13 densest pixels to mitigate artificial dipoles, resulting in a final catalog of 636,966 AGN candidates and leaving a sky fraction of 47.4 \%.

\section{Results} \label{sec:results}
\begin{figure*}[htbp]
    \centering
    \gridline{
        \fig{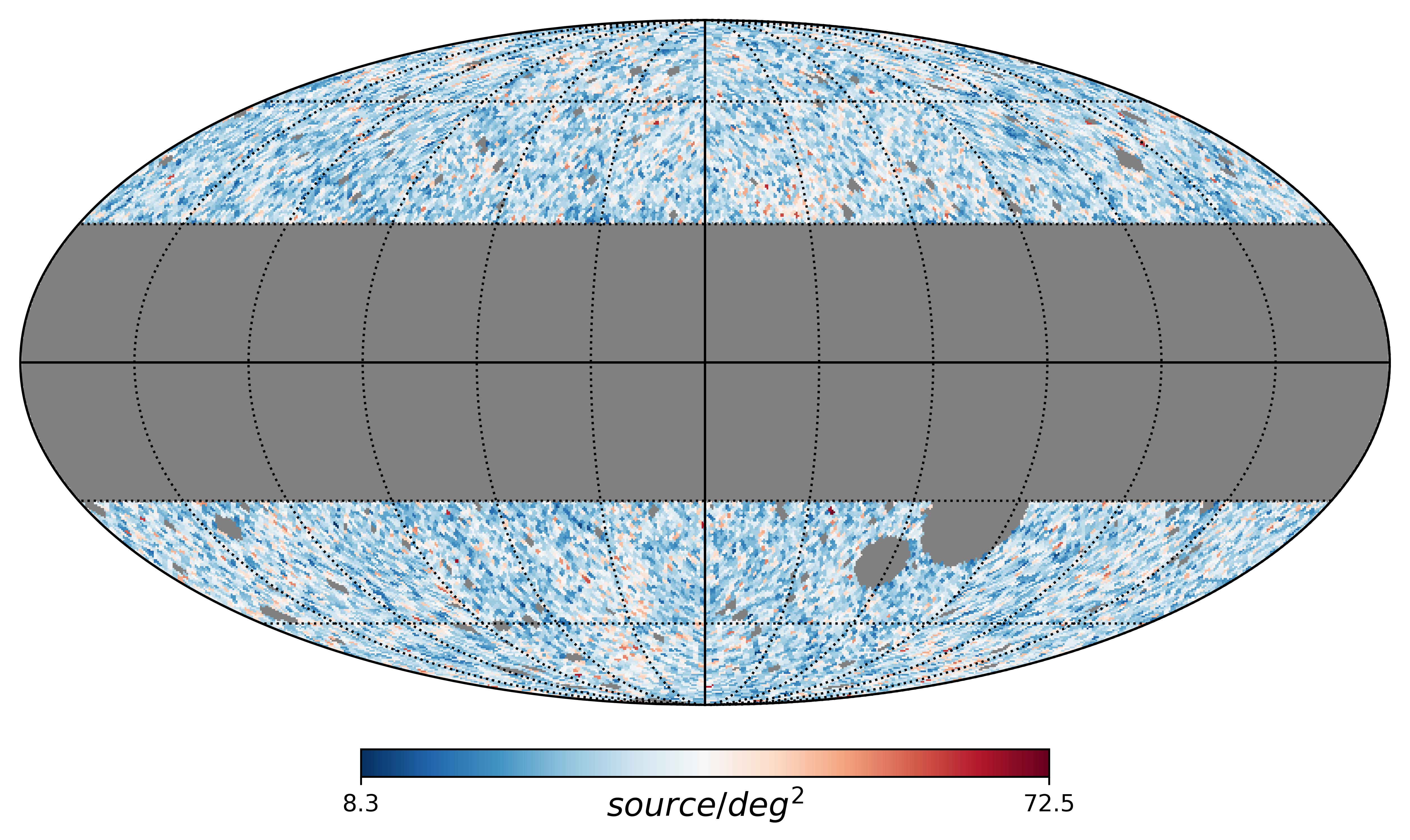}{0.49\textwidth}{(a)}{\label{fig:finalsamples}}
        \fig{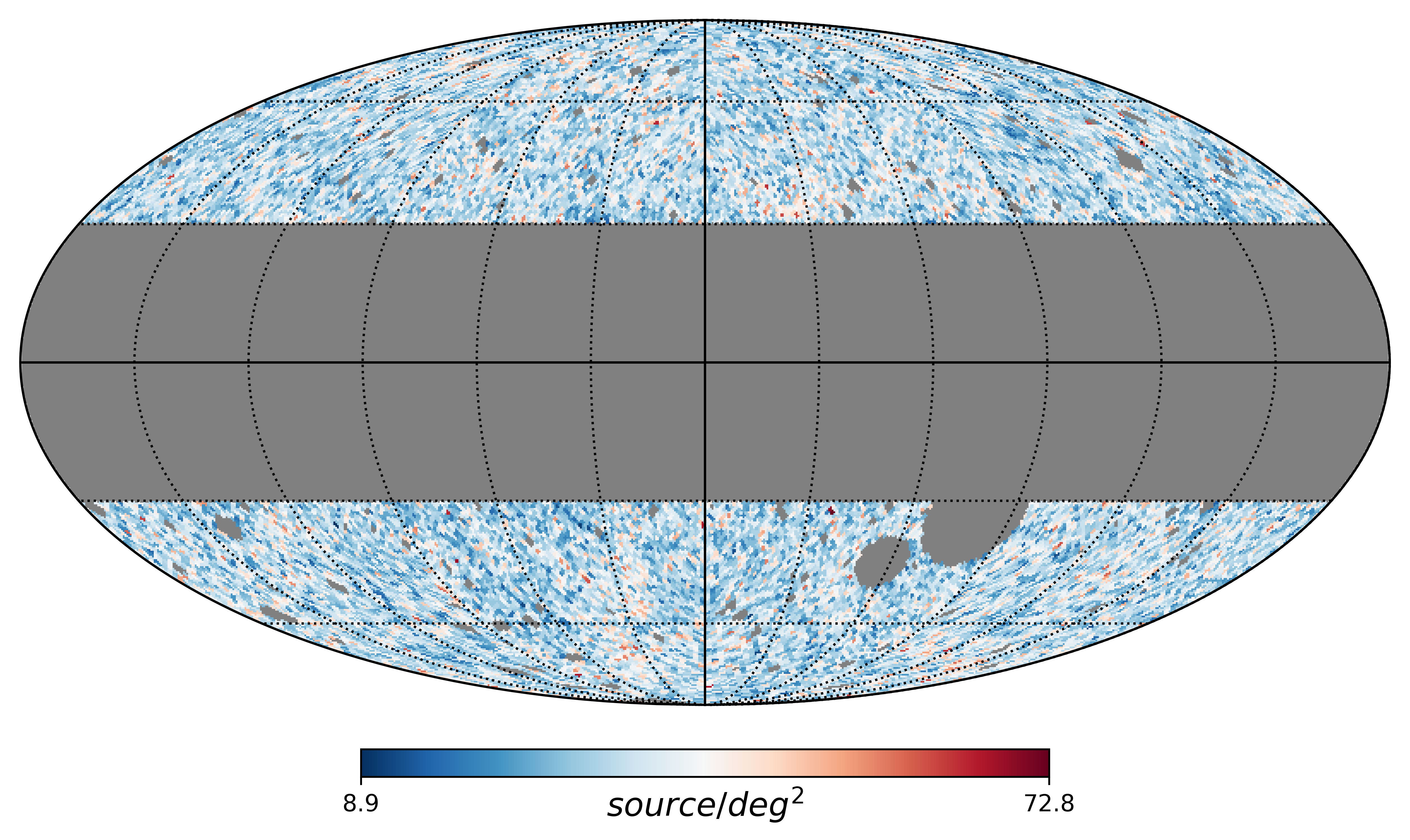}{0.49\textwidth}{(b)}{\label{fig:correction}}
             }
    \gridline{
        \fig{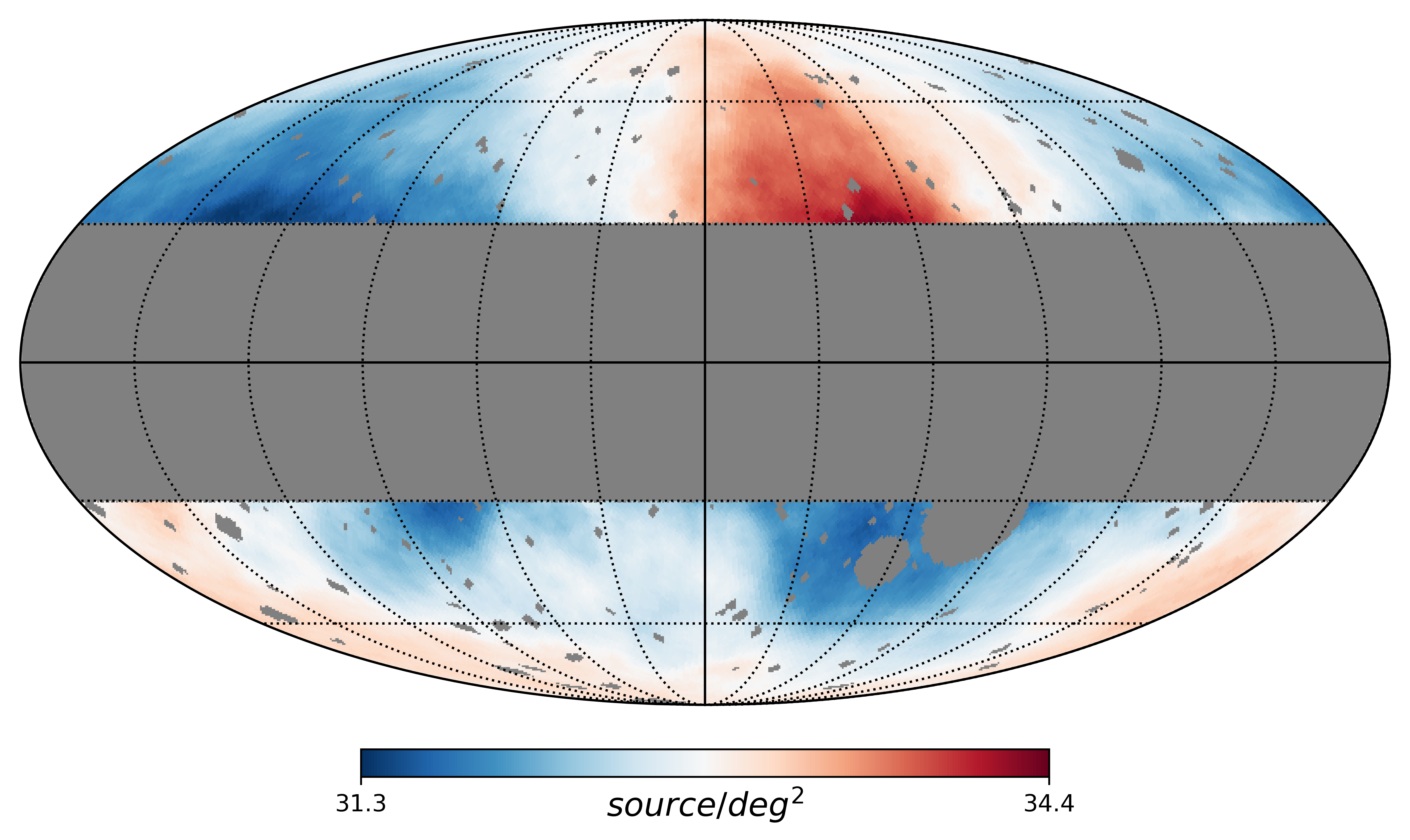}{0.49\textwidth}{(c)}{\label{fig:smoothing}}
        \fig{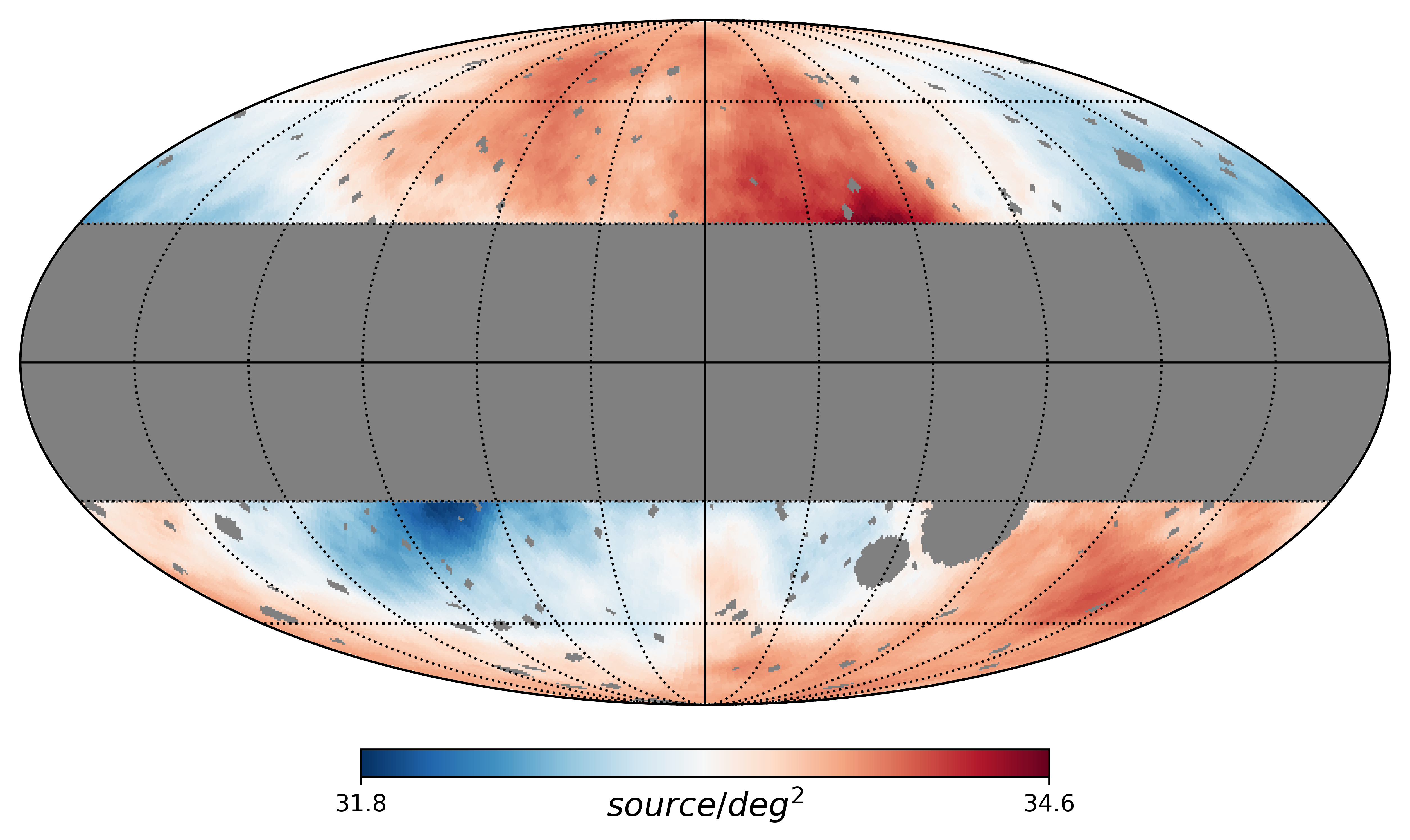}{0.49\textwidth}{(d)}{\label{fig:corsmoothing}}
             }
    \caption{Top: Number density maps of the final catalog before (a) and after (b) correcting for the ecliptic bias. Bottom: Smoothed number density maps before (c) and after (d) correcting for the ecliptic bias. {The unit of the number density is identical to that of Fig.~\ref{fig:AW_numden}.}}
    \label{fig:finalmap}
\end{figure*}

\begin{figure}[t!]
    \centering
    \includegraphics[width=0.95\linewidth]{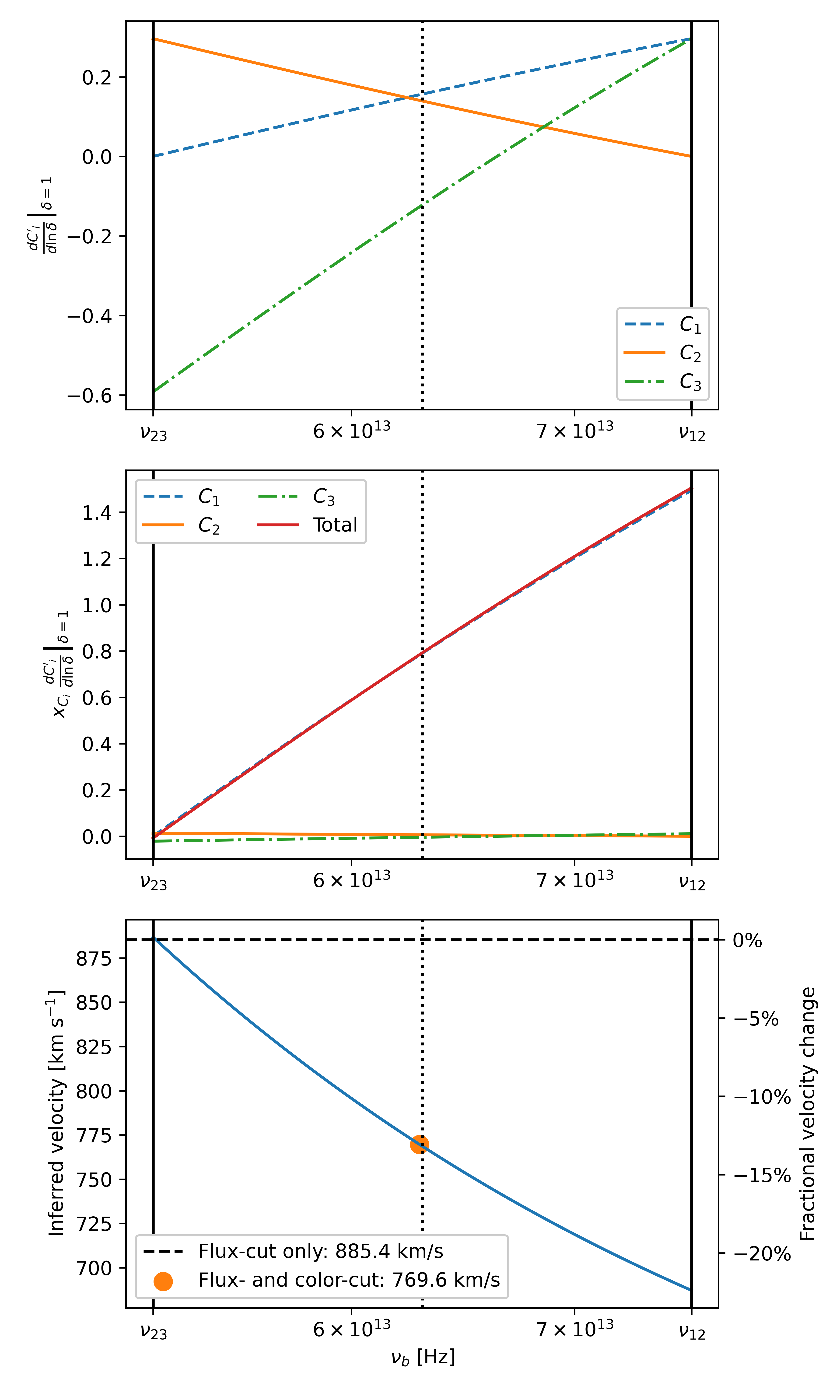}
    \caption{Dependence of the color-cut dipoles on the break frequency $\nu_b$. (Top) Derivatives of the color indices with respect to the log Doppler factor, $\left.dC'_i/d\ln\delta \right\vert_{\delta=1}$, as a function of the break frequency $\nu_{\rm b}$. The blue (dashed), orange (solid) and green (dot-dashed) lines correspond to $\left.d C'_{1}/d\ln\delta\right|_{\delta=1}$, $\left.d C'_{2}/d\ln\delta\right|_{\delta=1}$ and $\left.d C'_{3}/d\ln\delta\right|_{\delta=1}$, respectively.
    (Middle) The corresponding contributions to the kinematic dipole, $x_i \left.dC'_i/d\ln\delta \right\vert_{\delta=1}$ (Eqs.~\ref{eq:multicolordipole} and \ref{eq:finaleq}). The red curve represents their sum. This shows $C_1$ predominantly produces the color-cut dipole anisotropy.
    In both panels, the dotted vertical line marks the geometric mean of the two edges of the W2 band. See Appendix~\ref{app:colordipole} for the mathematical expression for $\left.d C'_{i}/d\ln\delta\right|_{\delta=1}$.
    (Bottom) Variance in the color-cut dipole anisotropy and the corresponding velocity estimate. For our fiducial case ($\nu_b =6.3\times 10^{13} \ \text{Hz}$, dotted line), marked by the orange circle, the color-cut dipole suppresses the velocity by $\sim 13\%$ from the value without the color-cut selection. For an allowed variation of $\nu_b$, the color-cut dipole anisotropy (blue solid line) suppresses the estimated velocity from the flux-cut only (dashed line) value, to as much as $\sim 22\%$.}
    \label{fig:Dc}
\end{figure}

As described in Sec.~\ref{sec:dipole}, the velocity equivalent of the measured dipole under the kinematic interpretation can be inferred from the observed dipole only after determining the spectral and population indices. Since the WISE magnitudes are given in the Vega system, we first convert them to the AB system \citep{Bessell2012}.
The offset between the AB and Vega systems, $m_{\text{AB}}=m_{\text{Vega}}+\text{offset}$, is 2.699, 3.339, 5.174, and 6.620~mag for the $W1$, $W2$, $W3$, and $W4$ bands, respectively. Using colors in the AB magnitude system, 
we construct lookup tables following \citet{Secrest2021} to derive the photometrically determined spectral power indices $\alpha_1$ and $\alpha_2$ ($S \propto \nu^{-\alpha}$), corresponding to the $[W1-W2]$ and $[W2-W3]$ bands, respectively. The distributions of these indices yield average power indices $\alpha_1=0.49$ and $\alpha_2=0.76$ for $C_1$ and $C_2$ color indices, respectively. For the population index $x$, defined as the slope of cumulative source counts above a flux threshold $\bar{S}_{W1}$, we obtain a mean value of $x=2.161$ when moving the flux threshold while the other color criteria are fixed. Applying the same procedure to the color criteria gives $x_{C_1}=5.044, \ x_{C_2}=0.044$, and $x_{C_3}=0.036$, respectively. This demonstrates that sample selection is primarily sensitive to $[W1-W2]$ color cut as can be seen in Fig.~\ref{fig:Dc}.

Fig.~\ref{fig:Dc} shows the color-selection contribution to the kinematic dipole as a function of the break frequency $\nu_{\rm b}$ within the $W2$ band. The left panel presents the derivatives $\left.dC'/d\ln\delta\right\vert_{\delta=1}$ for the three color cuts, while the right panel shows the corresponding products $x_{C_i} \left.dC'_i/d\ln\delta \right\vert_{\delta=1}$. The total contribution to the kinematic dipole, shown by the red curve, is determined predominantly by $i=1$ values and increases monotonically as $\nu_{\rm b}$ shifts from the lower- to upper-frequency edge of the $W2$ band. 

The kinematic dipole contributed by the color-cut criterion, or the new correction to the EB formula $\sum_{i}x_{C_i}\left.dC'_i/d\ln\delta \right\vert_{\delta=1}$, ranges from $-0.01$ to $1.49$. We estimate this contribution numerically by assuming a broken-power-law SED with a break frequency $\nu_{\rm b}$ located in the $W2$ band. In the left panel of Fig.~\ref{fig:Dc}, the color-cut-induced dipole components---namely, the unweighted derivatives of $C_1$, $C_2$, and $C_3$ with respect to $\ln \delta$---are shown as functions of the break frequency $\nu_{\rm b}$. The right panel presents the $x_{C_i}$-weighted color derivatives for the three cuts, together with the total color-cut-induced dipole component. In both panels, the vertical dotted line marks the geometric mean of the two edges of the $W2$ band, $\nu_{\rm b}=\sqrt{\nu'_{12}\nu'_{23}}$, where $\nu'_{23}=5.23 \times 10^{13} \ \text{Hz}$, $\nu'_{12}=7.59 \times 10^{13} \ \text{Hz}$. 

Our final catalog consists of 636,966 AGN candidates with a dipole amplitude of $\mathcal{D}=0.018$ in the direction of $(l=303.09^\circ, b=2.47^\circ)$. Implementing the measured spectral and population indices, we infer a velocity of $v=924.59$~km/s from the combined flux and color cuts, assuming that the observed dipole is purely kinematic, as described by Eq.~\ref{eq:colordipole}. For comparison, considering the flux cut contribution only yields $v=1063.67$~km/s. The observed dipole amplitude is 2.57 times as large as that of the CMB, located $56.65^\circ$ away from the direction of the CMB dipole. Fig.~\ref{fig:finalmap}(a) shows the number density map of the final AGN candidates. For visualization purposes only, we apply a smoothing scheme using a top-hat filter with a 1-steradian radius. The corresponding smoothed number density map is shown in Fig.~\ref{fig:finalmap}(c).


\begin{deluxetable}{lcc}
\tabletypesize{\footnotesize}
\tablewidth{0pt}
\tablecaption{{Summary of dipole estimation} \label{tab:summary}}
\tablehead{
\colhead{Parameter} & \colhead{\shortstack{\rule{0pt}{3ex}Ecliptic bias\\uncorrected}} & \colhead{\shortstack{\rule{0pt}{3ex}Ecliptic bias\\corrected}}
}
\startdata
Final source number & 636,966 & 651,537 \\
Sky fraction & \multicolumn{2}{c}{47.4\%} \\
CMB-predicted $\mathcal{D}$ & \multicolumn{2}{c}{0.007} \\
Dipole amplitude & 0.018 & 0.015 \\
Dipole direction $(l,b)$ & $(303.09^\circ, 2.47^\circ)$ & $(305.76^\circ, 5.58^\circ)$ \\
$\angle \mathrm{CMB}$ & $56.65^\circ$ & $55.39^\circ$ \\
Inferred velocity & $924.59$~km/s & $769.62$~km/s\\
\enddata
\end{deluxetable}

\subsection{Spatial systematics and robustness}
\label{subsec:robustness}
\begin{figure}[ht!]
    \centering
    \includegraphics[width=0.95\linewidth]{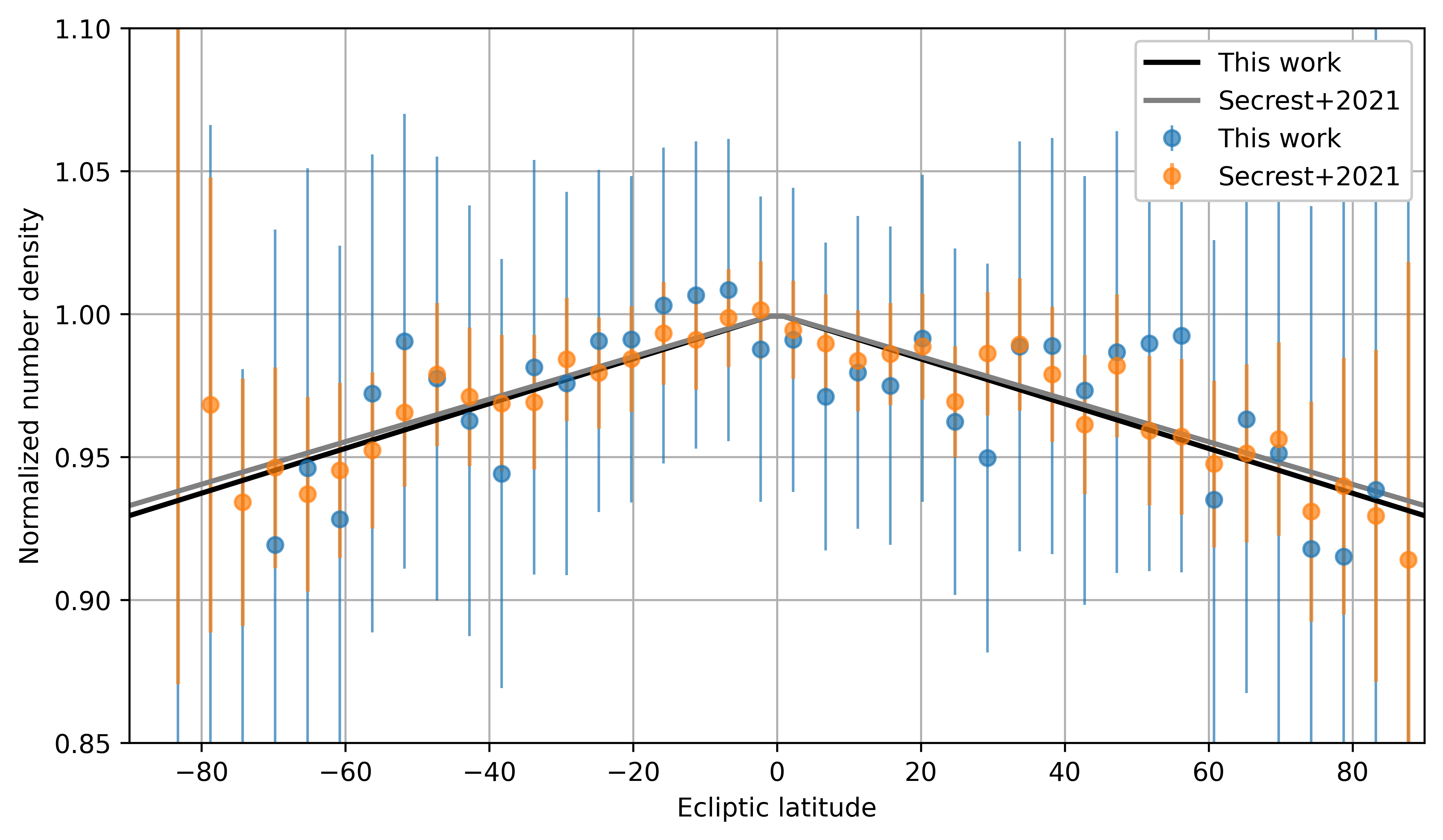}
    \caption{Ecliptic bias in AllWISE and CatWISE2020. The orange and blue error bars show the number densities and their corresponding uncertainties for AllWISE (limited by flux and 3-color cuts) and CatWISE2020. Both the number densities and their uncertainties are normalized to unity at the ecliptic plane for visualization purposes. The black and gray solid lines represent the linear trends for AllWISE and CatWISE2020, respectively.}
    \label{fig:eclbias}
\end{figure}
\citet{Secrest2021} reported an ecliptic bias in the ``CatWISE2020'' dipole analysis, in which $\mathcal{N}$ shows a linear dependence on ecliptic latitude ($\mathcal{N}$ increasing as the ecliptic latitude decreases). This trend is thought not to have a cosmological origin; instead, it can be partly accounted for by an anisotropic Eddington bias arising from the interplay between the anisotropic observational footprint and the photometric leakage of low-brightness sources into the high-brightness counts. \citep{Oayda2026}.

In our AllWISE AGN catalog, $\mathcal{N}$ bears an ecliptic-latitude dependence almost identical to that found in CatWISE2020, while remaining almost uniform in Galactic latitude.  We linearly fit these ecliptic biases in terms of the number density, obtaining a slope of $-0.025$ and an intercept of $32.45~\mathrm{deg}^{-2}$ at the ecliptic plane for our AllWISE AGN catalog, and a slope of $-0.051$ and an intercept of $68.89~\mathrm{deg}^{-2}$ for CatWISE2020. When normalized to unity at the ecliptic plane, the two number densities show ecliptic biases almost identical to each other (Fig.~\ref{fig:eclbias}). Using this ecliptic-latitude dependence of the AllWISE AGN map, we boost the number density at every HEALPix pixel and obtain a corrected number density map.

{To assess the sensitivity of the dipole estimation to the adopted ecliptic bias correction, we experiment with variance in the linear slope $m$ (the observed average density $\left<n\right>_{\rm obs}(\beta)\propto m\beta$ where $\beta$ is the ecliptic latitude). Across the tested settings with $\sim 1\sigma$ variation in $m$, the resulting variance in the dipole estimation does not alter our claim significantly. The deviation seen in this test lies in a few per cent level for the dipole amplitude and sub-degree level for the dipole direction (Table~\ref{tab:ecl_correction_sensitivity}). Therefore, under the limit of a linear regression and a $\beta$-only dependence, our dipole estimation is not too sensitive to the accuracy of the ecliptic bias correction.}


\begin{deluxetable}{lccccc}[ht!]
\tabletypesize{\footnotesize}
\tablecaption{{Sensitivity of the measured dipole to the empirical ecliptic-latitude correction.}\label{tab:ecl_correction_sensitivity}}
\tablehead{
  \colhead{} & 
  \colhead{$m$} & 
  \colhead{$D$} & 
  \colhead{$(l,b)$} & 
  \colhead{$\Delta\theta_{\mathrm{fid}}$} & 
  \colhead{$\angle \mathrm{CMB}$} 
}
\startdata
\shortstack[l]{\rule{0pt}{3ex}Uncorrected} & $0$ & $0.0185$ & $(303.09^\circ, 2.47^\circ)$ & $4.09^\circ$ & $56.65^\circ$ \\
\shortstack[l]{\rule{0pt}{3ex}Corrected} & $-0.025$ & $0.0154$ & $(305.76^\circ, 5.58^\circ)$ & $0$ & $55.39^\circ$ \\
\shortstack[l]{\rule{0pt}{3ex}Steep} & $-0.309$ & $0.0147$ & $(306.47^\circ, 6.40^\circ)$ & $1.08^\circ$ & $55.08^\circ$ \\
\shortstack[l]{\rule{0pt}{3ex}Shallow} & $-0.199$ & $0.0160$ & $(305.10^\circ, 4.82^\circ)$ & $1.00^\circ$ & $55.68^\circ$ \\
\enddata
\tablecomments{{
  The columns indicate the correction slope, the fractional dipole amplitude, Galactic direction, angular displacement from the fiducial corrected result, and separation from the adopted CMB dipole direction.}
}
\end{deluxetable}

With a total of 651,537 sources after bias correction, we obtain a dipole amplitude of $\mathcal{D}=0.015$, which is 2.14 times the dipole amplitude of the CMB. Assuming a purely kinematic origin for the observed dipole, the corresponding inferred velocities are $v=769.62$~km/s for the combined flux and color cuts and $v=885.39$~km/s for the flux cut alone. The direction of the bias-corrected dipole is pointing toward $(305.76^\circ, 5.58^\circ)$, with an angular separation of $55.39^\circ$ from the CMB dipole direction. We note that the photometric properties of the sources added by the bias correction are assumed to follow the distributions observed for the original AllWISE AGN candidates. The bias-corrected number density map and the corresponding smoothed density map are shown in Fig.~\ref{fig:finalmap}(b) and (d), respectively. The overall results before and after bias correction are summarized in Table~\ref{tab:summary}. 

Given the substantial discrepancy in the dipole direction compared with previous studies, we test the robustness of our analysis under variations in sample selection and masking, including the bright- and faint-end magnitude limits and the Galactic plane mask. We find that both the dipole amplitude and direction are relatively stable against changes to the bright-end magnitude limit, whereas the direction is more sensitive to the faint-end limit and Galactic mask. {The lowest faint-end limit and the small-latitude Galactic mask ($|b|<15^\circ$) we tested, however, should \emph{not} be taken as credible selection criteria and consequently, robustness with respect to the dipole direction remains satisfactory as well. First, bringing down the flux limit to fainter ends ($12<W1<15.4$) produces stripes of high density regions on the density map, clearly related to the anisotropic observational footprint. This tendency gets stronger as the limiting flux gets fainter. Second, the average number density at a given galactic latitude, $\left< n_{\rm obs}\right>(b)$, of our catalog is almost constant at $|b|\gtrsim30^\circ$ while precipitates quickly when $b$ gets below $20^\circ$. It is therefore not optimal to include all the sources at $|b|>15^\circ$. Excluding these two cases, the estimated dipole directions under our robustness test are roughly consistent with each other within $\lesssim 15^\circ$. } 

\subsection{Simulation}
\label{subsec:simulation}
\begin{figure}[htbp]
    \centering
    \gridline{\fig{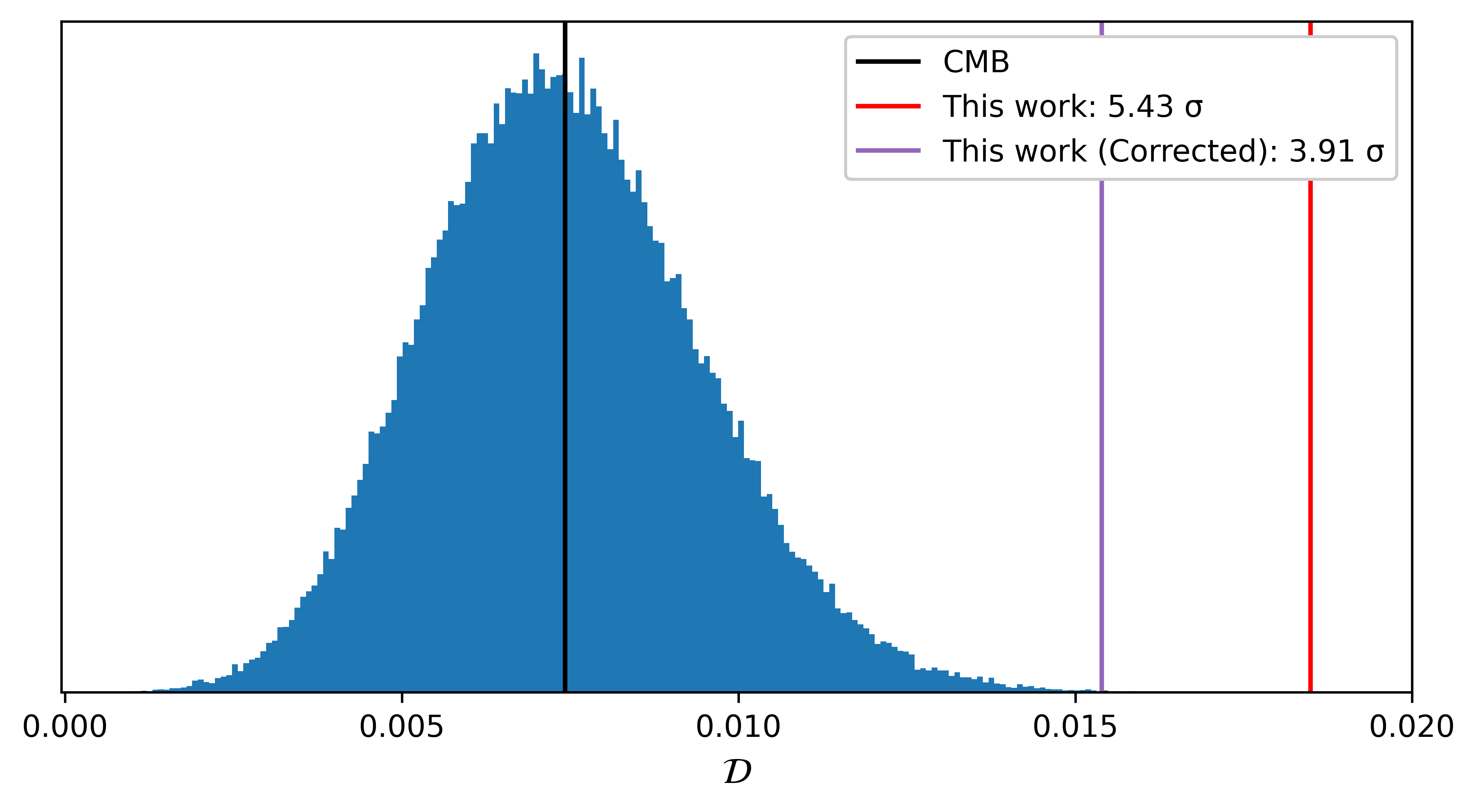}{0.45\textwidth}
                {\label{fig:amp}}}
    \gridline{\fig{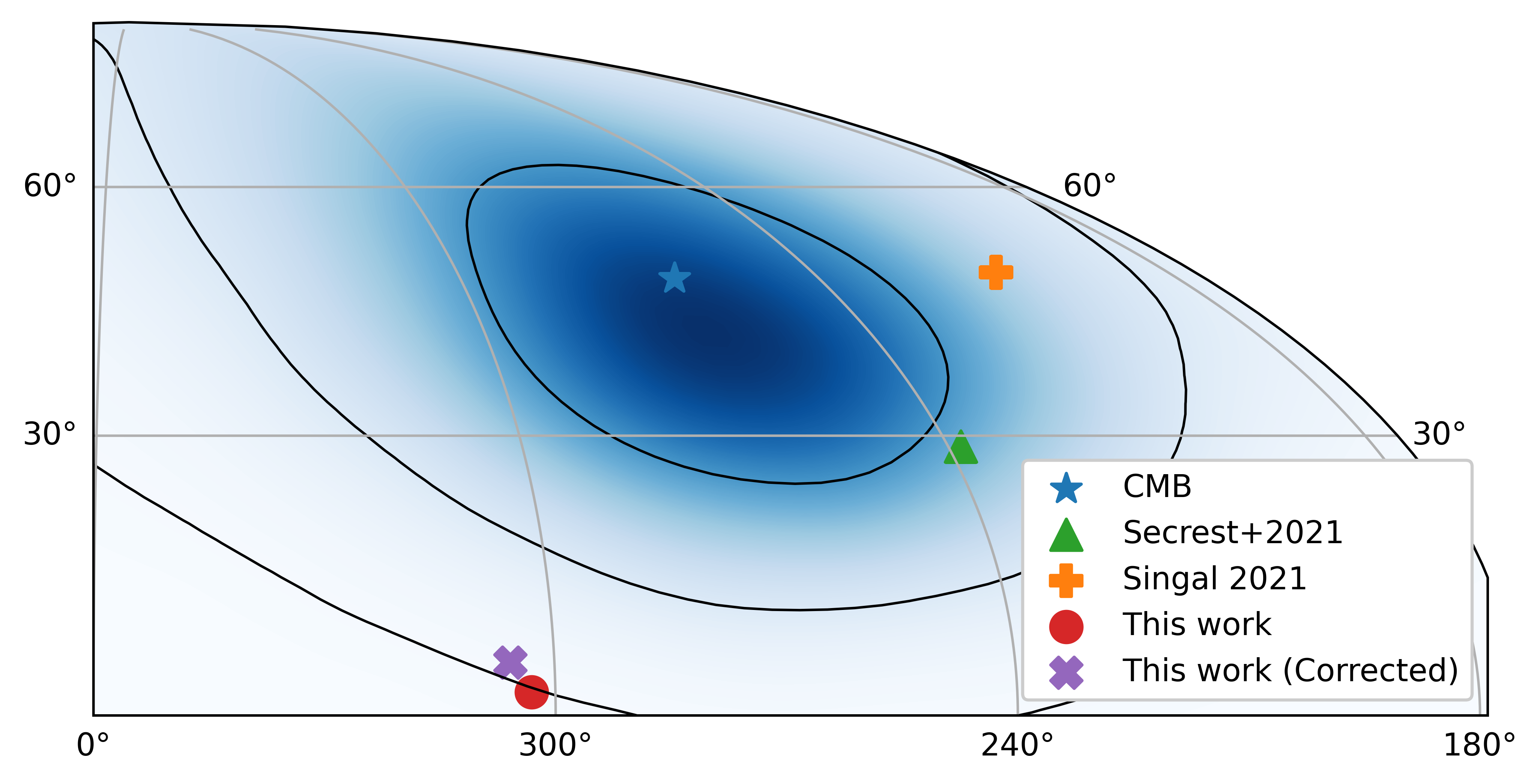}{0.45\textwidth}
                {\label{fig:dir}}}
    \caption{Top: Distribution of the simulated dipole amplitudes. The black, red, and purple solid lines represent the mean of the simulated dipole amplitudes, the dipole amplitude estimated from the AllWISE AGN candidates, and the dipole amplitude after the ecliptic bias correction, respectively.
    Bottom: Distribution of the simulated and estimated dipole directions in galactic coordinates. The probability density distribution of the simulated dipole directions is shown in pseudo-color (dark: dense, light: sparse), together with solid contour lines enclosing probabilities of 39.4\%, 86.5\%, and 98.9\% (corresponding to 2D-gaussian $1\sigma$, $2\sigma$, and $3\sigma$ levels, respectively). The directions obtained in this work (red circle and purple cross) lie along the outermost contour, corresponding to a tail probability of only $\sim 1.1\%$.
    Overlaid are the dipole directions of the CMB (blue, star), the CatWISE2020 QSO catalog by \citet[green, triangle]{Secrest2021}, the AllWISE AGN catalog by \citet[orange, +]{Singal2021}, the AllWISE AGN catalog in this work without (red, circle) and with (purple, x) a correction for the ecliptic bias. The directions obtained in this work lie along the outermost contour, corresponding to a tail probability of only $\sim 1.1\%$ to occur.}
    \label{fig:sim}
\end{figure}

To quantify how different the measured AGN dipole is from the CMB dipole in both amplitude and direction, we perform simulations. We developed a dipole simulation code and performed $10^5$ realizations in order to quantify the statistical significance of the estimated dipole, both in amplitude and direction, relative to the CMB dipole. In each iteration, several million sources were initially generated at random positions on the sky, and relativistic aberration was applied to them according to the velocity information (amplitude and direction) measured from the CMB dipole. Doppler boosting was then implemented using the distributions of flux, spectral index, and color indices in the AllWISE catalog, with these quantities sampled jointly to preserve their correlations. The dipole was measured for each realization after imposing the same flux cut, color cuts, and masking scheme used in the sample selection, followed by matching the size of the final AGN catalog made of $\sim 637,000$ sources.

The distribution of the simulated dipoles reflects the expected outcome for an AllWISE-like catalog whose kinematic dipole is set by the CMB dipole, while incorporating the same selection (flux-cut and color-cut) effects and masking used for the observed sample. In the top panel of Fig.~\ref{fig:sim}, the estimated dipole amplitude differs from the CMB-predicted amplitude by $5.43\sigma$, assuming a Gaussian distribution. When corrected by the ecliptic bias, the dipole amplitude is reduced from the uncorrected one, but still $3.91\sigma$ away from the CMB-predicted one (Fig.~\ref{fig:sim}). 
As for the dipole direction, as shown in Fig.~\ref{fig:sim}, our estimated direction lies near the boundary of the  $\sim98.9\%$ probability region for the dipole direction of a simulated map to fall near the direction of the CMB dipole (corresponding to the $3\sigma$-confidence contour for this roughly two-dimensional Gaussian distribution of simulated dipole directions). When the AllWISE AGN distribution map is corrected for the ecliptic bias, the dipole direction changes only slightly from the uncorrected one (Fig.~\ref{fig:sim}).
The angular separation between our measured direction and that reported by \citet{Secrest2021} is approximately $67^\circ$ for the uncorrected case, $68^\circ$ for the corrected case.

The dipole anisotropy of our AGN catalog appears to support the idea that the region of the Universe hosting these AGNs is very unusual, with intrinsic inhomogeneity that is much larger than in most other regions. It is because (1) the amplitude of the dipole is much larger than the CMB-predicted dipole, and (2) the direction of the dipole would be probabilistically unlikely ($\sim 1\%$ likely as in Fig.~\ref{fig:sim}) to be at the estimated location if the dipole were kinematically driven. This large discrepancy in the dipole direction is in contrast to some earlier studies by e.g. \citet{Secrest2021} and \citet{Singal2011} who find dipole directions marginally consistent with the CMB dipole. Our result is more natural because the discrepancy in the amplitude requires a physical (or systematic) origin different from the kinematic origin, and if so there is no reason why the dipole should be aligned with the kinematic dipole.

\section{Summary and Discussion}\label{sec:discussion}
We measured the large-scale number-count dipole, using the AGN candidates we selected from the AllWISE dataset, to test the consistency of the observed anisotropy with the kinematic expectation under the CP. In this work, we extended the Ellis-Baldwin formalism by adopting a broken-power-law SED, thereby incorporating the color-cut-induced dipole. The AGN candidates were selected by the 3-band criteria proposed by \citet{Blecha2018}, resulting in $\sim 6.4\times 10^5$ final AGN candidates. With the measured spectral and population indices, the inferred velocity under the kinematic interpretation from the AllWISE AGN catalog is $v=924.59$~km/s when both flux and color limits are taken into account. This value is $13.08\%$ lower than the velocity inferred in the flux-cut-only case, $v=1063.67$~km/s, demonstrating that the correction for color selection has a non-negligible impact on the inferred kinematic dipole. Despite this correction, the measured dipole amplitude remains discrepant from the CMB-predicted value at the $5.43\sigma$ level, {while the dipole direction deviates from the CMB-predicted direction by approximately $55^\circ$, corresponding to a $\sim 1.1\%$ tail probability under a two-dimensional Gaussian distribution.} The persistence of the dipole tension, together with the fact that the measured direction is {very different} from the direction reported by \citet{Secrest2021}, is noteworthy given the use of a newly compiled AllWISE AGN catalog and additional selection cuts designed to mitigate spatial selection effects.


Following \citet{EB84}, most dipole studies have adopted a single-power-law SED, which does not explicitly account for spectral breaks. To address this limitation, we adopted a more realistic, broken-power-law SED to describe the spectral behavior across the WISE bands. This allowed us to account for the additional dipole contribution arising from color selection. We focused primarily on the case in which $\nu_{\rm b}$ is located at the center of the $W2$ band. In addition, we also analyzed the derivatives of the colors as a function of $\nu_{\rm b}$ to investigate how the color-cut contribution depends on the location of $\nu_{\rm b}$ (Fig.~\ref{fig:Dc}), as our knowledge on $\nu_{\rm b}$ is limited by the photometric bandwidth. The total contribution from the color terms increases monotonically with increasing $\nu_{\rm b}$ across the $W2$ band. The minimum contribution occurs when $\nu_{\rm b}$ is at the lower-frequency edge of the $W2$ band, where the band-averaged flux is dominated by the shallower slope $\alpha_1$. On the other hand, the maximum contribution is obtained when $\nu_{\rm b}$ lies at the higher-frequency edge, where the steeper slope $\alpha_2$ becomes dominant. Owing to the large population index $x_{C_1}$, the color-cut-induced dipole is dominated by the $[W1-W2]$ cut among the three terms. 

Our inferred velocity from both limits is approximately 2.4 times larger than the velocity expected from the CMB dipole, indicating that a tension still remains. This tension is consistent with previous analyses based on the WISE catalogs: \citet{Singal2021} reported a velocity approximately four times larger than the CMB expectation using the AllWISE catalog, while \citet{Secrest2021} found a velocity more than twice the CMB expectation using the CatWISE2020 catalog. However, a detailed comparison should be made with caution because of differences in source selection criteria (e.g., color cuts and flux limits) and assumptions about the source spectra. Nonetheless, our results demonstrate that the kinematic dipole is sensitive not only to the flux limit but also to the color selection. The color-selection effects should therefore be carefully considered when interpreting the observed dipole as a kinematic velocity equivalent. 

{In contrast to the tension in the amplitude, the measured direction lies near the boundary of the $1 \%$ tail region,
whereas previous WISE-based studies found directions more consistent with the CMB dipole direction \citep{Singal2021, Secrest2021}, as shown in Fig~\ref{fig:sim}.} This appears to point to the following two causes.
(1) One possible explanation is that the inferred dipole direction depends on the adopted color-selection criteria. Although we obtain a dipole amplitude consistent with previous WISE-based studies, our use of a different color-selection criteria could have led to the discrepancy in the dipole direction. One can imagine that a new source catalog can be treated as a new ``realization'' of the dipole (see Section ``Simulation'') and thus the directional variance may occur. Our preliminary analyses using alternative AGN color cuts \citep{Jarrett2011, Mateos2012} suggest that the inferred dipole can indeed be sensitive to the adopted selection criteria, with amplitudes still similar to what we have reported here but directions varying from catalog to catalog.
A systematic investigation of this dependence will be presented in a forthcoming paper (Lee et al. in preparation).
(2) Another possible explanation is that local inhomogeneity provides a significant contribution to the observed dipole anisotropy. While rare \citep{Tiwari2016}, if this is the case then such a local contribution should not be aligned with the CMB dipole in general. The observed directional discrepancy thus highlights the need to consider contributions beyond a purely kinematic interpretation. Further study is warranted.


If the discrepancy cannot be fully explained by local and observational effects, a more fundamental possibility is that the matter-rest frame does not coincide with the CMB-rest frame. 
This possibility has been considered in previous studies, either by allowing for an intrinsic component of the observed CMB dipole \citep{Ma2011} or by constructing cosmological models with a relative bulk motion of matter with respect to the CMB-rest frame \citep{Martin2026}, even though the leverage permitting such models remains relatively limited. The possibility of a mismatch between the matter- and CMB-rest frames therefore seems slim, if not completely impossible. 

The SEDs of individual AGNs are more complex than the simple broken-power-law model used in this work. We assume that the sources are characterized by a common SED with representative values of spectral and population indices. In reality, AGN populations have different SEDs, including variations in spectral indices, spectral curvature, and the locations of spectral breaks. Incorporating individual and more realistic SED models in future analyses could reduce the systematic uncertainties in the inferred dipole. We also apply a mask to avoid contamination and artificial effects from nearby objects. This reduces the effective sky coverage available for estimating the dipole and may introduce additional systematic effects or biases. {We also find that source counts vary with ecliptic latitude and use a linear fit to correct for this trend. This empirical correction, however, does not capture the full complexity of the WISE survey selection function. 
Thus, the $3.91\sigma$ difference between our estimated dipole amplitude and the CMB dipole amplitude should be interpreted with caution.} 
Future analyses with improved masking schemes and more detailed modeling of spatial systematics will be important for assessing the robustness of the inferred dipole. 

Further investigation of the dipole anisotropy with independent source catalogs seems promising. For example, future observations with the Spectro-Photometer for the History of the Universe, Epoch of Reionization and Ices Explorer (SPHEREx) mission could provide a particularly valuable avenue as an independent probe. The large number of galaxies expected from SPHEREx, along with its strong capability to identify galaxies and AGNs, could substantially reduce the shot noise, while its spectral information could enable more reliable source identification than the color-based selection used in this work (see how LIM dipole emerges in \citealt{Ahn2025}). Extending the color-dipole formalism developed here to SPHEREx data and to kinematic dipole analyses of color-selected samples could provide new insights into the origin of the observed discrepancy between kinematic dipole measurements and the CMB dipole. 

\begin{acknowledgments}
KA is supported by the National Research Foundation of Korea (NRF) RS-2021-NR058956, RS-2025-16302968 and the Korea Astronomy and Space Science Institute under the R\&D program (Project No. 2025-9-844-00) supervised by the Korea AeroSpace Administration.
TTT is supported by JSPS Grant-in-Aid for Scientific Research (24H00247), JST CREST Grant No.\ JPMJCR24Q1, and by the Joint Research Program of the Institute of Statistical Mathematics (General Research 2), ``Machine-learning cosmogony: from structure formation to galaxy evolution.''
This publication makes use of data products from the Wide-field Infrared Survey Explorer, which is a joint project of the University of California, Los Angeles, and the Jet Propulsion Laboratory/California Institute of Technology, and NEOWISE, which is a project of the Jet Propulsion Laboratory/California Institute of Technology. WISE and NEOWISE are funded by the National Aeronautics and Space Administration.
\end{acknowledgments}

\begin{contribution}
D. Lee led the overall project, developed the theoretical framework, processed the survey data, and wrote the main text of the manuscript. K. Ahn co-led the project, developed the theoretical framework, and wrote a good part of the introduction. M. Oh and G. Asok contributed to the collection of the survey data, and validated the theoretical framework. T. T. Takeuchi contributed to the theoretical framework and provided scientific/technical feedback. All authors discussed the results and contributed to the final manuscript.



\end{contribution}

%



\appendix
\section{Modification of the Ellis \& Baldwin equation: Dipole anisotropy from color cuts}\label{app:colordipole}

\begin{figure}[ht!]
    \centering
    \gridline{\fig{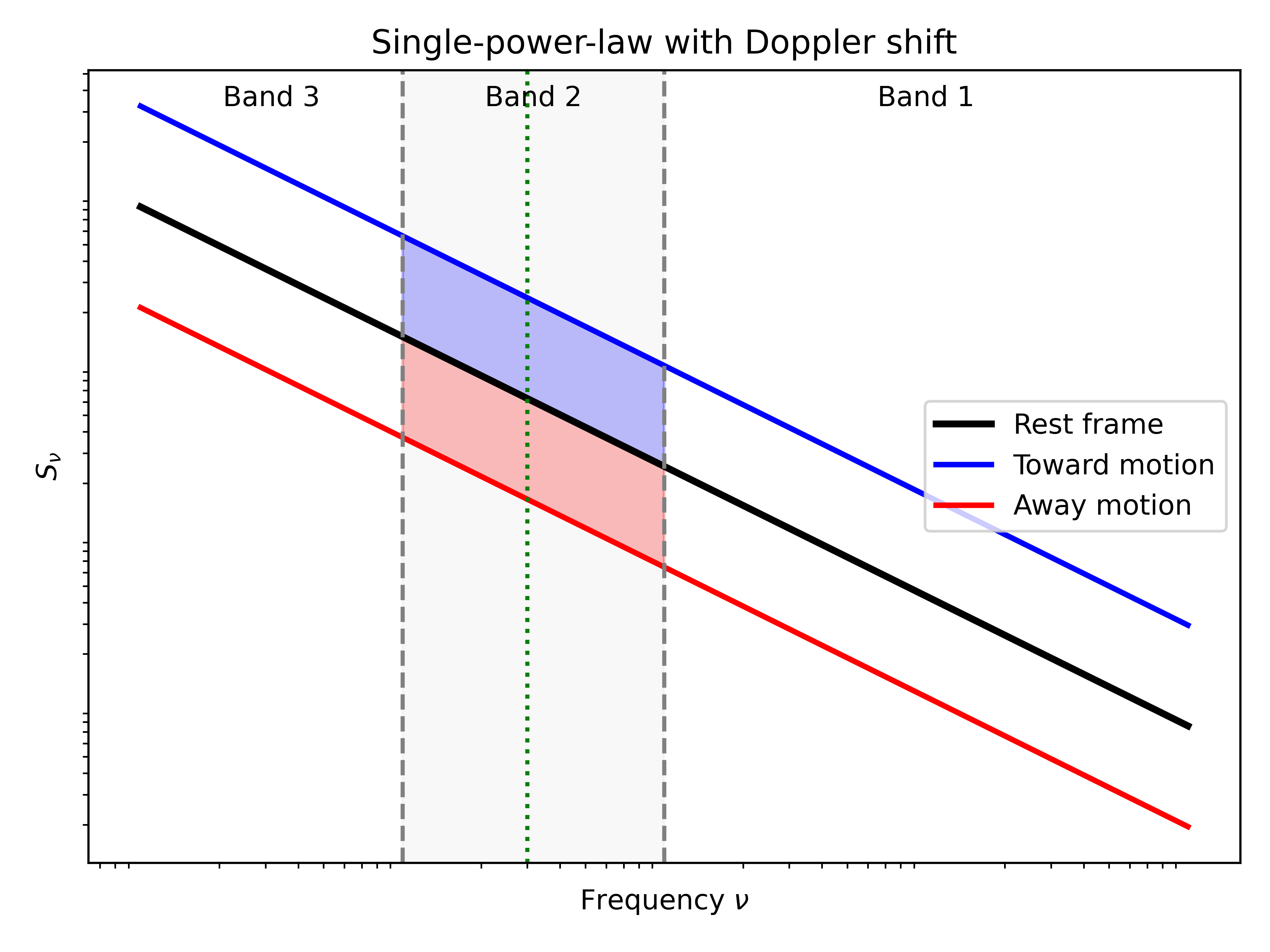}{0.49\textwidth}
                {(a) SPL}\label{fig:spl}
              \fig{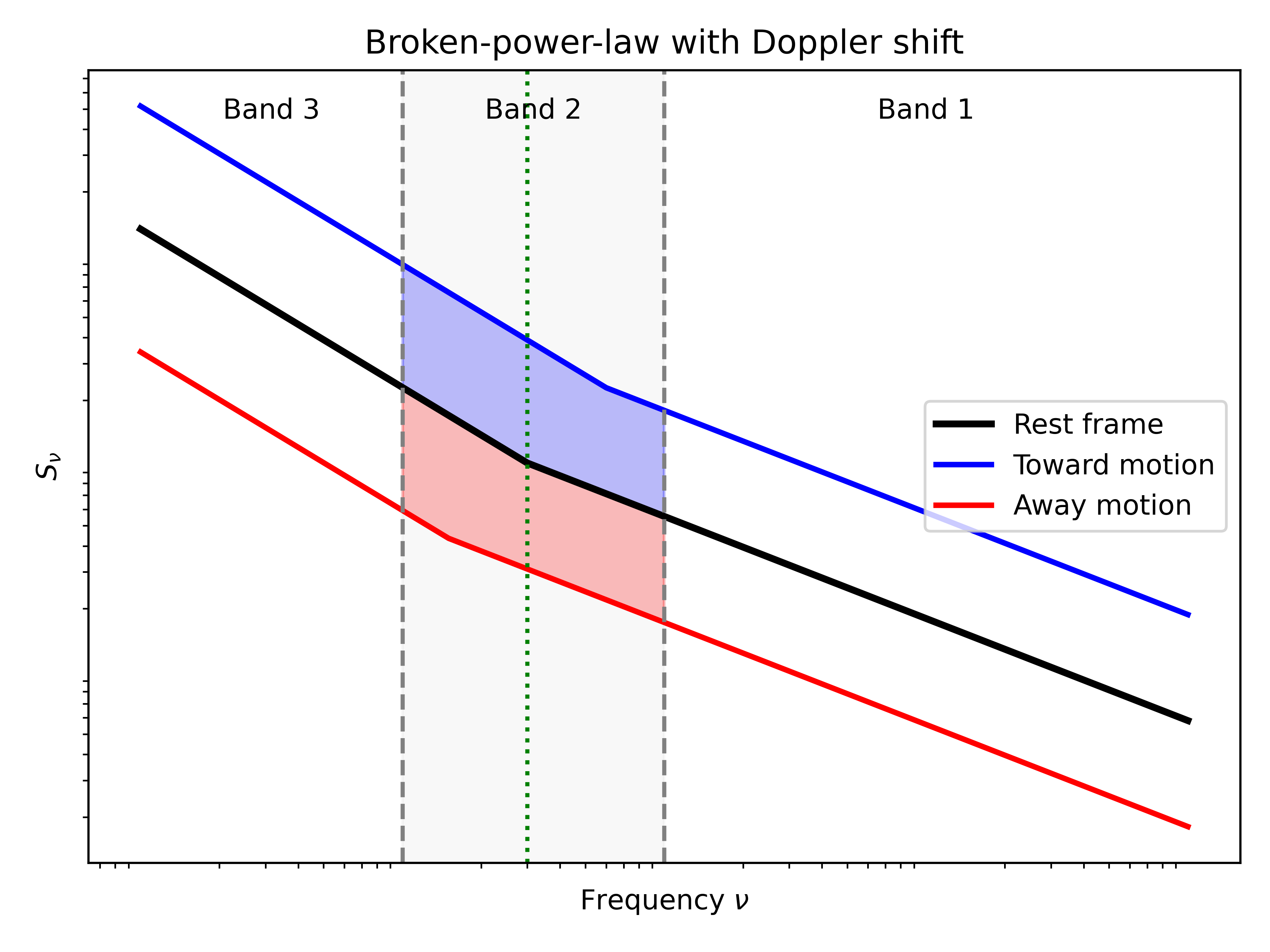}{0.49\textwidth}
                {(b) BPL}}\label{fig:bpl}
    \caption{Illustration of how radiation sources with a single-power-law SED (a) and a broken-power-law SED (b) is observed and produce a dipole anisotropy in observed colors in the latter case. (a) In a single-power-law SED case, the intrinsic color does not change thanks to the steadiness in spectral slopes regardless of the observer's motion. (b) On the contrary, in a broken-power-law case, a color dipole is generated due to change of the effective spectral slope due to the shift of the break frequency either to the right (blue) or to the left (red). }
    \label{fig:illust}
\end{figure}
Here we describe how a dipole anisotropy arises from a composite selection criterion made of a flux cut and multiple color cuts in detail. For notational convenience, we define the following evaluation-point notation: 
\begin{align*}
\left.A\right\vert_{0} \equiv \left.A\right\vert_{\delta=1}, \quad \left.A\right\vert_{\ell'} \equiv
\left.A\right\vert_{\bar{S}',\bar{C}'_1, \bar{C}'_2, \bar{C}'_3}.  
\end{align*}

In this work, we consider 3 color-cuts proposed by \citet{Blecha2018}, and assume a broken-power-law SED. Due to the spectral break, the colors become direction-dependent. Each color can then be expanded to first order in $\ln\delta$, 
\begin{eqnarray}
    C'_1(\delta) &\approx& C'_1(\delta=1) + \left.\frac{dC'_1}{d\ln\delta}\right\vert_{0} \ln\delta = C_1 + \left.\frac{dC_1'}{d\ln\delta}\right\vert_{0} \ln\delta, \nonumber \\
    C'_2(\delta) &\approx& C'_2(\delta=1) + \left.\frac{dC'_2}{d\ln\delta}\right\vert_{0} \ln\delta = C_2 + \left.\frac{dC_2'}{d\ln\delta}\right\vert_{0} \ln\delta, \nonumber \\
    C'_3(\delta) &\approx& C'_3(\delta=1) + \left.\frac{dC'_3}{d\ln\delta}\right\vert_{0} \ln\delta = C_3 + \left.\frac{dC_3'}{d\ln\delta}\right\vert_{0} \ln\delta.
\end{eqnarray}
If the observed colors exceed the limiting values, $\bar{C}_1$, $\bar{C}_2$, and $\bar{C}_3$, the corresponding colors in the rest frame are
\begin{align}
    C_1 > \bar{C}_1 - \left.\frac{dC'_1}{d\ln\delta}\right\vert_{0} \ln\delta, \quad
    C_2 > \bar{C}_2 - \left.\frac{dC'_2}{d\ln\delta}\right\vert_{0} \ln\delta, \quad
    C_3 > \bar{C}_3 - \left.\frac{dC'_3}{d\ln\delta}\right\vert_{0} \ln\delta.
\end{align}

We can define the differential number density as
\begin{eqnarray}
    n(S, C_{1}, C_{2}) = \frac{dN}{dS \ dC_{1} \ dC_{2} \ d\Omega},
\end{eqnarray}
and the cumulative number density with 1 flux cut and 3 color cuts
\begin{align}
    \mathcal{N}(>\bar{S},>\bar{C}_{1}, >\bar{C}_{2}, >\bar{C}_{3}) &= \frac{dN}{d\Omega}(>\bar{S}, >\bar{C}_{1}, >\bar{C}_{2}, >\bar{C}_{3}) \nonumber \\
                                     &= \int_{\bar{S}} dS \int_{\bar{A}} dC_{1} dC_{2}\ n(S,C_{1}, C_{2}),
\end{align}
where the integral over color indices is to represent the 2-dimensional integral only within a region $\bar{A}$ enclosed by the color-cut limits seen in Fig.~\ref{fig:colorcut}. 
Lagrangian number is conserved in the two frames, $dN'=dN$, and thus we have
\begin{align}
    \mathcal{N'}(>\bar{S}',>\bar{C}'_1, >\bar{C}'_2, >\bar{C}'_3) &= \mathcal{N}(>\bar{S},>\bar{C}_1, >\bar{C}_2, >\bar{C}_3) \frac{d\Omega}{d\Omega'} \nonumber \\
                                         &= \mathcal{N}(>\bar{S},>\bar{C}_1, >\bar{C}_2, >\bar{C}_3)\delta^2.
\label{eq:finaleq}
\end{align}
A first-order Taylor expansion of $\mathcal{N}(>\bar{S},>\bar{C}_1, >\bar{C}_2, >\bar{C}_3)$ about ORF variables $\bar{S}',\bar{C}'_1, \bar{C}'_2, \bar{C}'_3$ gives

\begin{align}
    &\mathcal{N}(>\bar{S},>\bar{C}_1, >\bar{C}_2, >\bar{C}_3) \nonumber \\
    &\quad \approx \mathcal{N}(>\bar{S}',>\bar{C}'_1, >\bar{C}'_2, >\bar{C}_3') + \left. \frac{\partial \mathcal{N}}{\partial \bar{S}} \right\vert_{\ell'} (\bar{S}-\bar{S}')+ \left. \frac{\partial \mathcal{N}}{\partial \bar{C}_1} \right\vert_{\ell'} (\bar{C}_{1}-\bar{C}'_{1}) +\left. \frac{\partial \mathcal{N}}{\partial \bar{C}_{2}} \right\vert_{\ell'} (\bar{C}_{2}-\bar{C}'_{2}) +\left. \frac{\partial \mathcal{N}}{\partial \bar{C}_{3}} \right\vert_{\ell'} (\bar{C}_{3}-\bar{C}'_{3}) \nonumber \\
    &\quad = \mathcal{N}(>\bar{S}',>\bar{C}'_{1},>\bar{C}'_{2}, >\bar{C}_{3}') \nonumber \\
    &\quad \quad \times \left\{1+(\delta^{-(1+\alpha)}-1) \left. \frac{\partial \ln\mathcal{N}}{\partial \ln \bar{S}}\right\vert_{\ell'} -\left.\frac{\partial \ln\mathcal{N}}{\partial \bar{C}_{1}}\right\vert_{\ell'}\left.\frac{dC'_{1}}{d\ln\delta}\right\vert_{0} \ln\delta -\left.\frac{\partial \ln\mathcal{N}}{\partial \bar{C}_{2}}\right\vert_{\ell'}\left.\frac{dC'_{2}}{d\ln\delta}\right\vert_{0} \ln\delta -\left.\frac{\partial \ln\mathcal{N}}{\partial \bar{C}_{3}}\right\vert_{\ell'}\left.\frac{dC'_{3}}{d\ln\delta}\right\vert_{0} \ln\delta \right\}.
\end{align}
Expanding to first order in $\beta$, 
\begin{align}
    &\mathcal{N'}(>\bar{S}',>\bar{C}'_{1}, >\bar{C}'_{2}, >\bar{C}_{3}')  \nonumber \\
    &\quad = \mathcal{N}(>\bar{S},>\bar{C}_{1}, >\bar{C}_{2}, >\bar{C}_{3}') \delta^2 \nonumber \\
    &\quad = \mathcal{N}(>\bar{S}',>\bar{C}'_{1}, >\bar{C}'_{2}, >\bar{C}_{3}')\left\{1 + 2\beta\cos\theta -(1+\alpha) \left. \frac{\partial \ln\mathcal{N}}{\partial \ln \bar{S}}\right\vert_{\ell'} \beta\cos\theta \right.  \nonumber \\
    &\quad \quad  \left. -\left.\frac{\partial \ln\mathcal{N}}{\partial \bar{C}_{1}}\right\vert_{\ell'}\left.\frac{dC'_{1}}{d\ln\delta}\right\vert_{0} \beta\cos\theta -\left.\frac{\partial \ln\mathcal{N}}{\partial \bar{C}_{2}}\right\vert_{\ell'}\left.\frac{dC'_{2}}{d\ln\delta}\right\vert_{0} \beta\cos\theta -\left.\frac{\partial \ln\mathcal{N}}{\partial \bar{C}_{3}}\right\vert_{\ell'}\left.\frac{dC'_{3}}{d\ln\delta}\right\vert_{0} \beta\cos\theta \right\}.
\end{align}
Finally, the dipole moment of $\mathcal{N}$ is given by
\begin{align}
    &\mathcal{N'}_\text{dip}(>\bar{S}',>\bar{C}'_{1}, >\bar{C}'_{2}, >\bar{C}_{3}') \nonumber \\
    &\quad = \mathcal{N}(>\bar{S}',>\bar{C}'_{1}, >\bar{C}'_{2}, >\bar{C}_{3}') \left\{ 2+x(1+\alpha)  +x_{C_1}\left.\frac{dC'_{1}}{d\ln\delta}\right\vert_{0} +x_{C_2}\left.\frac{dC'_{2}}{d\ln\delta}\right\vert_{0} +x_{C_3} \left.\frac{dC'_{3}}{d\ln\delta}\right\vert_{0} \right\}\beta,\label{aeq:dipmo}
\end{align}
where $x=-\left.\partial \ln\mathcal{N} / \partial \ln \bar{S}\right\vert_{\ell'}$, $x_{C_1} = -\left.\partial \ln\mathcal{N} / \partial \bar{C}_{1}\right\vert_{\ell'}$, $x_{C_2} = -\left.\partial \ln\mathcal{N} / \partial \bar{C}_{2}\right\vert_{\ell'}$, and $x_{C_3} = -\left.\partial \ln\mathcal{N} / \partial \bar{C}_{3}\right\vert_{\ell'}$.

To numerically estimate the color-cut-induced dipole, we adopt a broken power-law SED as a toy model for the average SED of AllWISE AGNs. This is to roughly encapsulate the deviation of the general SED from the single power-law SED while limited by the spectral ambiguity from photometry. The broken-power-law SED has a spectral break at $\nu_{\rm b}$ located within the W2 band (band 2 in Fig.~\ref{fig:illust}), to yield
\begin{eqnarray}
    S(\nu) =  
    \begin{cases}
        S_b \left(\nu/\nu_{\rm b}\right)^{-\alpha_{2}} & \textrm{if} \quad \nu<\nu_{\rm b} \\
        S_b \left(\nu/\nu_{\rm b}\right)^{-\alpha_{1}} & \textrm{if} \quad \nu \ge\nu_{\rm b}.
    \end{cases}
\end{eqnarray}
Here, $\alpha_1$ and $\alpha_2$ are given in Sec~\ref{sec:results}, with $\alpha_1=0.49$ and $\alpha_2=0.76$. We require the break frequency to lie within $\nu'_{23} < \nu_{\rm b} < \nu'_{12}$, where $\nu'_{23}=5.23 \times 10^{13}$ Hz and $\nu'_{12}=7.59 \times 10^{13}$ Hz. Large bandwidth in photometry inhibits one from fixing $\nu_{\rm b}$. The numerical behavior of the color-cut contribution for this toy model is shown in Fig.~\ref{fig:Dc}. The general solution (Eq.~\eqref{aeq:dipmo}) can be written explicitly in this toy model with parameters for the broken power-law: 
\begin{align}
    &\left.\frac{dC'_{1}}{d\ln\delta}\right\vert_{0} = -B\left((1+\alpha_1) - \frac{\frac{1}{\alpha_2}\left[(1+\alpha_2)\left(\frac{\nu'_{23}}{\nu_b}\right)^{-\alpha_2}-1\right]-\frac{1}{\alpha_1}\left[(1+\alpha_1)\left(\frac{\nu'_{12}}{\nu_b}\right)^{-\alpha_1}-1\right]}{\frac{1}{\alpha_2}\left[\left(\frac{\nu'_{23}}{\nu_b}\right)^{-\alpha_2}-1\right]-\frac{1}{\alpha_1}\left[\left(\frac{\nu'_{12}}{\nu_b}\right)^{-\alpha_1}-1\right]}\right), \nonumber \\
    &\left.\frac{dC'_{2}}{d\ln\delta}\right\vert_{0} = -B\left(\frac{\frac{1}{\alpha_2}\left[(1+\alpha_2)\left(\frac{\nu'_{23}}{\nu_b}\right)^{-\alpha_2}-1\right]-\frac{1}{\alpha_1}\left[(1+\alpha_1)\left(\frac{\nu'_{12}}{\nu_b}\right)^{-\alpha_1}-1\right]}{\frac{1}{\alpha_2}\left[\left(\frac{\nu'_{23}}{\nu_b}\right)^{-\alpha_2}-1\right]-\frac{1}{\alpha_1}\left[\left(\frac{\nu'_{12}}{\nu_b}\right)^{-\alpha_1}-1\right]} - (1+\alpha_2)\right), \nonumber \\
    &\left.\frac{dC'_{3}}{d\ln\delta}\right\vert_{0} = -B\left((1+\alpha_1) + 2(1+\alpha_2)- 3\frac{\frac{1}{\alpha_2}\left[(1+\alpha_2)\left(\frac{\nu'_{23}}{\nu_b}\right)^{-\alpha_2}-1\right]-\frac{1}{\alpha_1}\left[(1+\alpha_1)\left(\frac{\nu'_{12}}{\nu_b}\right)^{-\alpha_1}-1\right]}{\frac{1}{\alpha_2}\left[\left(\frac{\nu'_{23}}{\nu_b}\right)^{-\alpha_2}-1\right]-\frac{1}{\alpha_1}\left[\left(\frac{\nu'_{12}}{\nu_b}\right)^{-\alpha_1}-1\right]}\right),
\end{align}
where $B\equiv2.5/\ln10$ and the other parameters appearing on the right-hand sides are described in Sec.~\ref{sec:results}.

For given photometric data, there exists a spectral uncertainty. Therefore we also test a running power-law SED, as an alternative SED, to calculate the color-cut dipole. We take a simplistic form for the SED, given by
\begin{equation}
    S(\nu) = S_b\left(\frac{\nu}{\nu_b}\right)^{-\alpha+\frac{1}{2}\eta \ln(\nu/\nu_b)}.
\end{equation}
Here, $\nu_b$ is a reference frequency at which $S_b$ is defined. If we enforce the condition that $S(\nu)$ runs through the three points ($\nu_1$, $S(\nu_1)$), ($\nu_2$, $S(\nu_2)$) and ($\nu_3$, $S(\nu_3)$), the parameters $\alpha$ and $\eta$ are related to the two spectral indices as
\begin{equation}
    \alpha = \alpha_{1}+\frac{\eta}{2}\ln\left(\frac{\nu_1\nu_2}{\nu^2_b}\right)
    = \alpha_{2}+\frac{\eta}{2}\ln\left(\frac{\nu_2\nu_3}{\nu^2_b}\right), \quad
    \eta = \frac{2(\alpha_{1}-\alpha_{2})}{\ln (\nu_3/\nu_1)},
\end{equation}
where $\nu_i$ denotes the geometric mean frequency of the corresponding WISE band, with $\nu_1=9.48 \times 10^{13}$ Hz, $\nu_2=6.30\times 10^{13}$ Hz, and $\nu_3=2.34 \times 10^{13}$ Hz. For $\eta>0$, the general solution (Eq.~\eqref{aeq:dipmo}) can be written explicitly for the running power-law:
\begin{align}
    &\left.\frac{dC'_{12}}{d\ln\delta}\right\vert_{0} = B \sqrt{\frac{2\eta}{\pi}} \left(\frac{\exp[z'^2_{1u}]-\exp[z'^2_{1d}]}{\operatorname{erfi}(z'_{1u})-\operatorname{erfi}(z'_{1d})} - \frac{\exp[z'^2_{2u}]-\exp[z'^2_{2d}]}{\operatorname{erfi}(z'_{2u})-\operatorname{erfi}(z'_{2d})}\right), \nonumber \\
    &\left.\frac{dC'_{23}}{d\ln\delta}\right\vert_{0} = B \sqrt{\frac{2\eta}{\pi}} \left(\frac{\exp[z'^2_{2u}]-\exp[z'^2_{2d}]}{\operatorname{erfi}(z'_{2u})-\operatorname{erfi}(z'_{2d})} - \frac{\exp[z'^2_{3u}]-\exp[z'^2_{3d}]}{\operatorname{erfi}(z'_{3u})-\operatorname{erfi}(z'_{3d})}\right), \nonumber \\
    &\left.\frac{dC'_{123}}{d\ln\delta}\right\vert_{0} = B \sqrt{\frac{2\eta}{\pi}} \left(\frac{\exp[z'^2_{1u}]-\exp[z'^2_{1d}]}{\operatorname{erfi}(z'_{1u})-\operatorname{erfi}(z'_{1d})} + 2\frac{\exp[z'^2_{3u}]-\exp[z'^2_{3d}]}{\operatorname{erfi}(z'_{3u})-\operatorname{erfi}(z'_{3d})} - 3\frac{\exp[z'^2_{2u}]-\exp[z'^2_{2d}]}{\operatorname{erfi}(z'_{2u})-\operatorname{erfi}(z'_{2d})}\right), 
\end{align}
where
\begin{equation}
    z'_i = \sqrt{\frac{\eta}{2}}\left( \ln\left(\frac{\nu'_i}{\nu_b}\right)-\frac{\alpha}{\eta}\right),
\end{equation}
and the subscripts $u$, $d$ indicate the upper and lower frequency limits of each band, respectively. 
Applying this running power-law model, including color-cut dipole anisotropy, yields a 14\% lower inferred velocity than the flux-cut only case, almost identical with the result obtained using the broken power-law model yielding a 13\% lower inferred velocity. Therefore, unless the actual AGN SED deviates significantly from our assumed toy models, the expected color-cut dipole impact will remain at $\lesssim 20\%$ level (see Fig.~\ref{fig:Dc}).

\section{Robustness test}\label{app:rob}
\begin{deluxetable}{lcccc}[htbp]
\tablewidth{0pt}
\tablecaption{Robustness test results of the dipole estimation under different selection criteria and masks \label{tab:robust}}
\tablehead{
  \colhead{Case} & 
  \colhead{Source Number} & 
  \colhead{Dipole Amplitude} & 
  \colhead{Dipole Direction} & 
  \colhead{$\angle \mathrm{CMB}$} \\
  \colhead{} & 
  \colhead{\footnotesize [Corrected]} & 
  \colhead{\footnotesize [Corrected]} & 
  \colhead{\footnotesize [Corrected]} & 
  \colhead{\footnotesize [Corrected]}
}
\startdata
\rule{0pt}{2.5ex}This work & 636,966 & 0.018 & $(303.09^\circ, 2.47^\circ)$ & $56.65^\circ$ \\
\footnotesize (Mask 30, $12 < W1 < 15.2$) & \footnotesize [651,537] & \footnotesize [0.015] & \footnotesize [$(305.76^\circ, 5.58^\circ)$] & \footnotesize [$55.39^\circ$] \\[4pt]
\hline
\rule{0pt}{2.5ex}Mask 15 & 928,673 & 0.022 & $(247.60^\circ, 4.47^\circ)$ & $46.00^\circ$ \\
& \footnotesize [954,389] & \footnotesize [0.020] & \footnotesize [$(243.26^\circ, 7.58^\circ)$] & \footnotesize [$44.34^\circ$] \\[4pt]
\hline
\rule{0pt}{2.5ex}$12 < W1 < 15.4$ & 940,606 & 0.018 & $(280.99^\circ, 2.72^\circ)$ & $47.77^\circ$ \\
& \footnotesize [966,221] & \footnotesize [0.015] & \footnotesize [$(279.24^\circ, 6.66^\circ)$] & \footnotesize [$43.51^\circ$] \\[4pt]
\hline
\rule{0pt}{2.5ex}$12 < W1 < 15.0$ & 422,947 & 0.023 & $(330.73^\circ, 2.12^\circ)$ & $73.02^\circ$ \\
& \footnotesize [431,064] & \footnotesize [0.021] & \footnotesize [$(335.11^\circ, 3.93^\circ)$] & \footnotesize [$74.47^\circ$] \\[4pt]
\hline
\rule{0pt}{2.5ex}$12 < W1 < 14.8$ & 278,266 & 0.026 & $(334.34^\circ, -0.55^\circ)$ & $77.38^\circ$ \\
& \footnotesize [283,279] & \footnotesize [0.024] & \footnotesize [$(338.15^\circ, 0.75^\circ)$] & \footnotesize [$78.86^\circ$] \\[4pt]
\hline
\rule{0pt}{2.5ex}$11.8 < W1 < 15.2$ & 637,561 & 0.019 & $(303.27^\circ, 2.25^\circ)$ & $56.94^\circ$ \\
& \footnotesize [652,192] & \footnotesize [0.016] & \footnotesize [$(305.95^\circ, 5.29^\circ)$] & \footnotesize [$55.73^\circ$] \\[4pt]
\hline
\rule{0pt}{2.5ex}$12.2 < W1 < 15.2$ & 636,193 & 0.018 & $(303.26^\circ, 2.45^\circ)$ & $56.76^\circ$ \\
& \footnotesize [650,795] & \footnotesize [0.015] & \footnotesize [$(306.01^\circ, 5.60^\circ)$] & \footnotesize [$55.50^\circ$] \\
\enddata
\tablecomments{The columns indicate the final source number, measured dipole amplitude, direction, and the angular separation from the direction of the CMB dipole. Quantities without and with square brackets represent the corresponding cases without and with the ecliptic bias correction, respectively.}
\end{deluxetable}

\begin{figure}[htbp]
    \centering
    \gridline{\fig{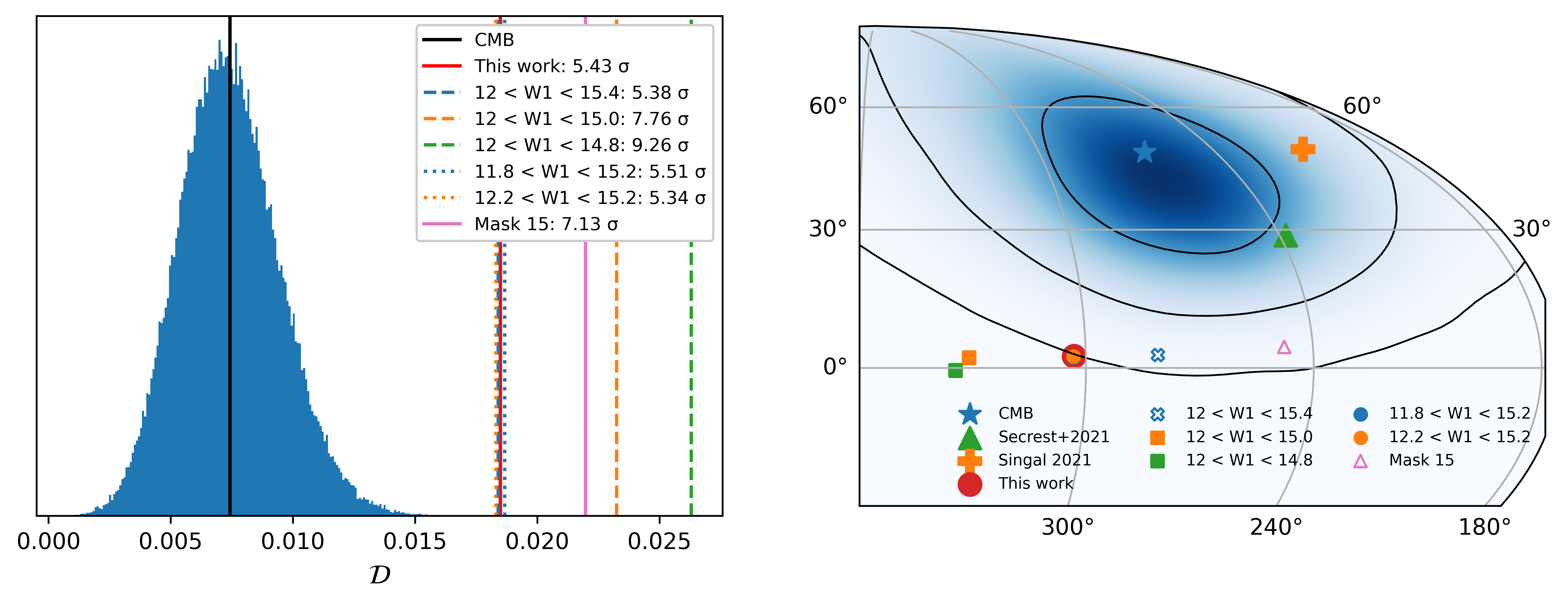}{0.99\textwidth}{(a)}
                {\label{}}}
    \gridline{\fig{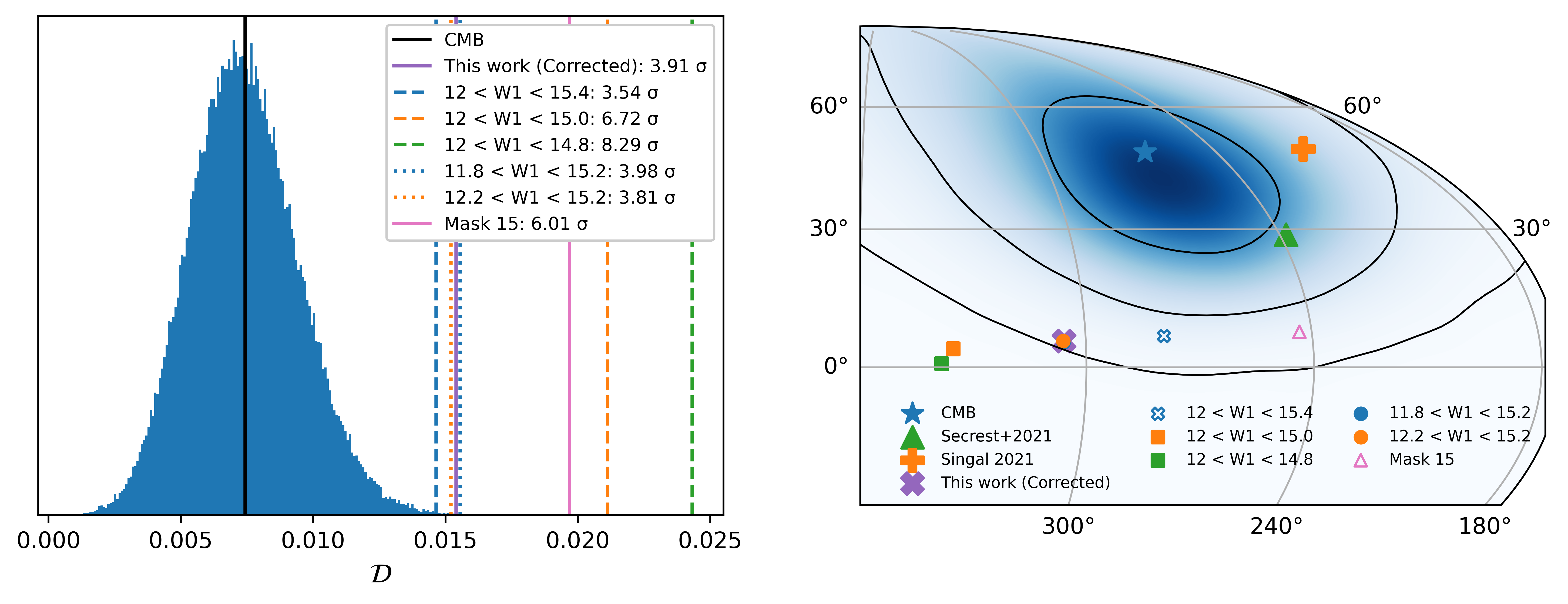}{0.99\textwidth}{(b)}
                {\label{}}}
    \caption{Robustness tests for the cases without (a) and with (b) ecliptic bias correction. Left: Distribution of the simulated dipole amplitudes, as in the top panel of Fig~\ref{fig:sim}, with the robustness-test results. The dashed, dotted, and pink solid lines indicate the variations in the faint- and bright-end limits of the W1 magnitude cut and the sky mask, respectively. Right: Distribution of the simulated dipole directions in Galactic coordinates, as in the bottom panel of Fig~\ref{fig:sim}, with the robustness-test results. The {open cross, filled square, filled circle, and open triangle} markers indicate variations in the faint- and bright-end limits of the W1 magnitude cut and the sky mask, respectively. {{Note that open symbols ($12<W1<15.4$ and Mask 15) denote cases suffering strong systematic effects (Section~\ref{subsec:robustness}) and thus should not be taken as credible estimations.}} 
    }
    \label{fig:robust}
\end{figure}
The substantial mismatch we find in the direction of the dipole compared with earlier studies necessitates a robustness test. Therefore, we examine how variations in the sample selection and masking affect the dipole estimation, both to assess the robustness of our results and to justify our analyses.
We first vary the faint-end limit of the W1 magnitude cut to 15.4, 15.0, and 14.8, yielding source counts of 940,606 (966,221), 422,947 (431,064), and 278,266 (283,279) for the uncorrected (corrected) cases, respectively. The dipole amplitudes for these cuts are shown as dashed lines in the left panels of Fig.~\ref{fig:robust}, while the corresponding dipole directions are indicated by {open cross and filled square} markers in the right panels. The top and bottom panels show the results for the uncorrected and corrected cases, respectively. For the higher faint-end limit {(open cross)}, $12 < W1 < 15.4$, the dipole amplitude is similar to what we obtained in this work, while the direction is closer to that of the CMB. Despite this, as described in Sec.~\ref{sec:sample}, we adopt an additional flux cut because of the stripe-like pattern observed in the number density map. Even with the magnitude cut $12 < W1 < 15.4$, weak stripe-like structures remain, potentially introducing an additional artificial dipole. 
In contrast, for the lower faint-end limits {(filled square)}, $12 < W1 < 15.0$ and $12 < W1 < 14.8$, the dipole amplitudes are much higher and the dipole directions deviate further from our result, although the directions appear to converge. However, these cases have smaller sample sizes and are therefore more susceptible to shot noise.
We then vary the bright-end limit of the W1 magnitude cut to 11.8 and 12.2, obtaining source counts of 637,561 (652,192) and 636,193 (650,795), respectively. In Fig.~\ref{fig:robust}, the dotted lines in the left panels represent the dipole amplitudes, while the {filled circle} markers in the right panels indicate the corresponding dipole directions. Changing the bright-end limit has a negligible effect on the dipole analysis, with both the amplitude and direction remaining consistent with our result, since relatively few sources are affected by this cut.

We also measure the dipole using an alternative Galactic plane mask with a masking threshold of $|b|<15^\circ$, while keeping all other conditions unchanged. The resulting dipole amplitudes are shown as pink solid lines in the left panels of Fig.~\ref{fig:robust}, and the corresponding directions are indicated by pink {open triangles} in the right panels. The dipole directions are closer to those reported for the CMB and in previous WISE-based studies, despite higher dipole amplitudes. However, the number density exhibits a clear decreasing trend as a function of absolute Galactic latitude around $|b| \sim 15^\circ$--$30^\circ$. As this trend could lead to an additional systematic effect, we adopt a $|b| <30^\circ$ mask for our analysis. All cases are summarized in Table~\ref{tab:robust}. {We note to the reader that the cases with open symbols should not be taken too seriously, as we explain their unwanted data corruption in Section~\ref{subsec:robustness}.}



\bibliography{refs_rev}{}
\bibliographystyle{aasjournalv7}



\end{document}